\documentclass[10pt,journal,compsoc]{IEEEtran}
\ifCLASSOPTIONcompsoc
  \usepackage[nocompress]{cite}
\else
  \usepackage{cite}
\fi

\usepackage{enumitem}
\usepackage{comment}
\usepackage{float}
\usepackage{graphics}
\usepackage{epstopdf}

\usepackage[linesnumbered,ruled,vlined,boxed,commentsnumbered]{algorithm2e}
\usepackage{tikz}
\newcommand*\circled[1]{\tikz[baseline=(char.base)]{\node[shape=circle,draw,inner sep=.5pt] (char) {#1};}}
\usepackage{amsmath}
\usepackage{algorithmic}

\begin{document}

%
\title{Efficient Time-Evolving Stream\\ Processing at Scale}
%
%
%
%
\author{Yu~Huang
\IEEEcompsocitemizethanks{\IEEEcompsocthanksitem Y. Huang is with Services Computing Technology and System Lab, Big Data Technology and System Lab, Cluster and Grid Computing Lab, School of Computer Science and Technology, Huazhong University of Science and Technology, Wuhan, 430074, China.
\protect\\
E-mail: yuh@hust.edu.cn
}
}
\IEEEtitleabstractindextext{%
\begin{abstract}

Time-evolving stream datasets exist ubiquitously in many real-world applications where their inherent hot keys  often  evolve  over times. Nevertheless, few existing solutions can provide efficient load balance on these time-evolving datasets while preserving low memory overhead.
In this paper, we present a novel grouping approach (named FISH), which can provide the efficient time-evolving stream processing at scale. The key insight of this work is that the keys of time-evolving stream data can have a skewed distribution within any bounded distance of time interval. This enables to accurately identify the recent hot keys for the real-time load balance within a bounded scope. We therefore propose an {\em epoch-based} recent hot key identification with specialized intra-epoch frequency counting (for maintaining low memory overhead) and inter-epoch hotness decaying (for suppressing superfluous computation). We also propose to heuristically {\em infer} the accurate information of remote workers through computation rather than communication for cost-efficient worker assignment.
We have integrated our approach into Apache Storm.
Our results on a cluster of 128 nodes for both synthetic and real-world stream datasets show that FISH significantly outperforms state-of-the-art with the average and the 99th percentile latency reduction by 87.12\% and 76.34\% (vs. W-Choices), and memory overhead reduction by 99.96\% (vs. Shuffle Grouping).


\end{abstract}

}

\maketitle

\IEEEdisplaynontitleabstractindextext

%
\IEEEpeerreviewmaketitle

\IEEEraisesectionheading{\section{Introduction}\label{sec:introduction}}

\IEEEPARstart {S}{treaming} processing has an important role in solving many real-world problems. From fraud detection (e.g., real-time financial activity~\cite{parikh2008scalable}) to real-time recommendations (e.g., analytics over microblogs~\cite{wang2016efficient,sharma2016graphjet} and live streaming~\cite{liao}), applications that generate stream data are ubiquitous.
Unlike structured stream data in which hot keys are relatively evenly distributed during the whole lifetime~\cite{Wikipedia}, real-world stream datasets often exhibit the unique feature that their inherent hot keys often evolve over times. One key is hot in some interval may be non-hot in the next interval. A typical example includes {\sf twitter} dataset where its catchword may vary frequently for different instants of time. At present, it also becomes greatly necessary and important to efficiently process these {\em time-evolving} stream datasets.

Reasonably distributing time-evolving stream datasets on a cluster of machines can provide the beneficial businesses with the cost-effective services. In an effort to exploit maximum benefits, time-evolving stream processing systems need to do the best at two aspects at least. First, all loads for time-evolving stream datasets must be balanced to a maximum extent. This indicates whether each worker is fully mobilized. It also directly affects the overall latency and throughput of stream processing. Second, considering the state of the stream data is backed up on multiple workers, the combined memory overhead on all machines should be controlled with a minimum of duplicates. This indicates how much memory is stored redundantly, which directly influences the scalability of stream processing systems.


Unfortunately, few existing solutions can meet all these two hard requirements. Fields Grouping utilizes key-based routing, which is prone to load imbalance across multiple workers~\cite{Storm}. Shuffle Grouping~\cite{Storm} uses round-robin manner to assign the loads. However, it potentially replicates the states associated with keys on each worker with a linear proportion of the memory overhead.
Other solutions attempt to balance the loads by leveraging operator migration~\cite{shah2003flux,xing2005dynamic,gedik2014partitioning,gedik2014elastic,castro2013integrating,chen2016bufferbank,basanta2017patterns}. A part of the keys are allowed to be rebalanced when load imbalance is detected. A number of studies~\cite{nasir2015power, nasir2016two} also aim to reduce the rebalancing overhead by identifying hot key and further assigning more workers. These earlier efforts on structured stream processing make a significant progress on getting a reasonable tradeoff, which, however, is far unsatisfactory from practical use for time-evolving stream processing. This is particularly true when the number of machines is scaled (as discussed in Section 2.3). In this paper, we are addressing whether and how we can build such an efficient and scalable time-evolving stream processing system. 

Nevertheless, there remains tremendously challenging to build a time-evolving stream processing system with all the desired properties satisfied.
First, since time-evolving stream processing often involves a large number of recent hot key identification operations, it should be not only accurate but also efficient, which is notoriously difficult.
In order to track most recently-occurred keys, it necessarily has to preserve a large amount of key-related information. Although existing approaches make a great progress on accuracy, the expense is that a substantial amount of computation~\cite{mabroukeh2010taxonomy,shan2009frequent,lim2014fast} or memory overhead~\cite{golab2003identifying,chang2003finding,arasu2004approximate,wang2005tfp,deng2012new} has occurred.

Second, handling time-evolving stream dataset may also need a timely adjustment for load balance at every moment, which is also difficult. Even worse, heterogeneous resources may further exacerbate this problem. To assign the appropriate workers for load balance, the servers have to frequently collect the state information from workers with considerable communication overhead~\cite{pietzuch2006network,buddhika2017online}. It remains challenging to make an efficient decision of worker assignment for preserving the real-time load balance.

In this paper, we propose an efficient grouping approach (named FISH) to process time-evolving streaming data at scale. Interestingly, we observe that, no matter how large a time interval is, the keys of time-evolving stream dataset within this bounded scope have a skewed power-law distribution where a small fraction of keys dominate most loads. This therefore allows to achieve real-time load balance within a bounded time interval by using hierarchical treatment~\cite{todtling2005one}. We present an epoch-based approach to accurately identify recent hot key. Each {\em epoch} can be a custom-sized key sequence. Intra-epoch identification counts the occurrence number of the key, which only stores the number of most frequent keys~\cite{manku2002approximate,karger2004simple} for preserving the low memory overhead. Inter-epoch identification uses time-aware approach~\cite{mabroukeh2010taxonomy,shan2009frequent,lim2014fast}, which adopts epoch-level (rather than tuple-level) update for reducing the superfluous computation.
To ensure the efficiency of worker assignment, we also recognize the simplicity of operations and the similarity of keys. We further propose a heuristic approach to infer (rather than prohibitively communicate) the information of remote worker in a more efficient manner.

This paper makes the following contributions:

\begin{itemize}
    \item We make a comprehensive study on the deficiencies of state-of-the-art grouping schemes for time-evolving stream datasets in terms of load balance and scalability.
    \item We present an efficient and scalable grouping scheme with epoch-based hot key identification and heuristic worker assignment, which can provide low-latency and high-throughput time-evolving stream processing.
    \item We evaluate our approach on both synthetic and real-world stream datasets. Experimental results show that our approach significantly outperforms state-of-the-art with the average and 99th percentile latency reduction by 87.12\% and 76.34\% (vs. W-Choices), and 96.66\% memory consumption reduction (vs. Shuffle Grouping).

\end{itemize}

The rest of this paper is organized as follows. We first give the background and motivation in  Section~\ref{sec:Background and Motivation}. Section~\ref{sec:Overview} provides the overview of our approach. Section~\ref{sec:Mechanism} elaborates the design of FISH. Section~\ref{sec:Extension} describes the extension for handling dynamic scenario with worker variation. Section~\ref{sec:Evaluation} discusses the results. We survey the related work in Section~\ref{sec:Related work} and conclude this work in Section~\ref{sec:Conclusion}.

\section{Background and Motivation}\label{sec:Background and Motivation}
In this section, we first briefly review the background of distributed stream processing and existing stream partitioning schemes. We next investigate the potential inefficiency of existing solutions towards time-evolving stream dataset through a comprehensive motivating study, finally followed by several challenges for coping with the problem.

\subsection{Distributed Stream Processing}
\begin{figure}[t]
  \centering
  \includegraphics[width=3.5in]{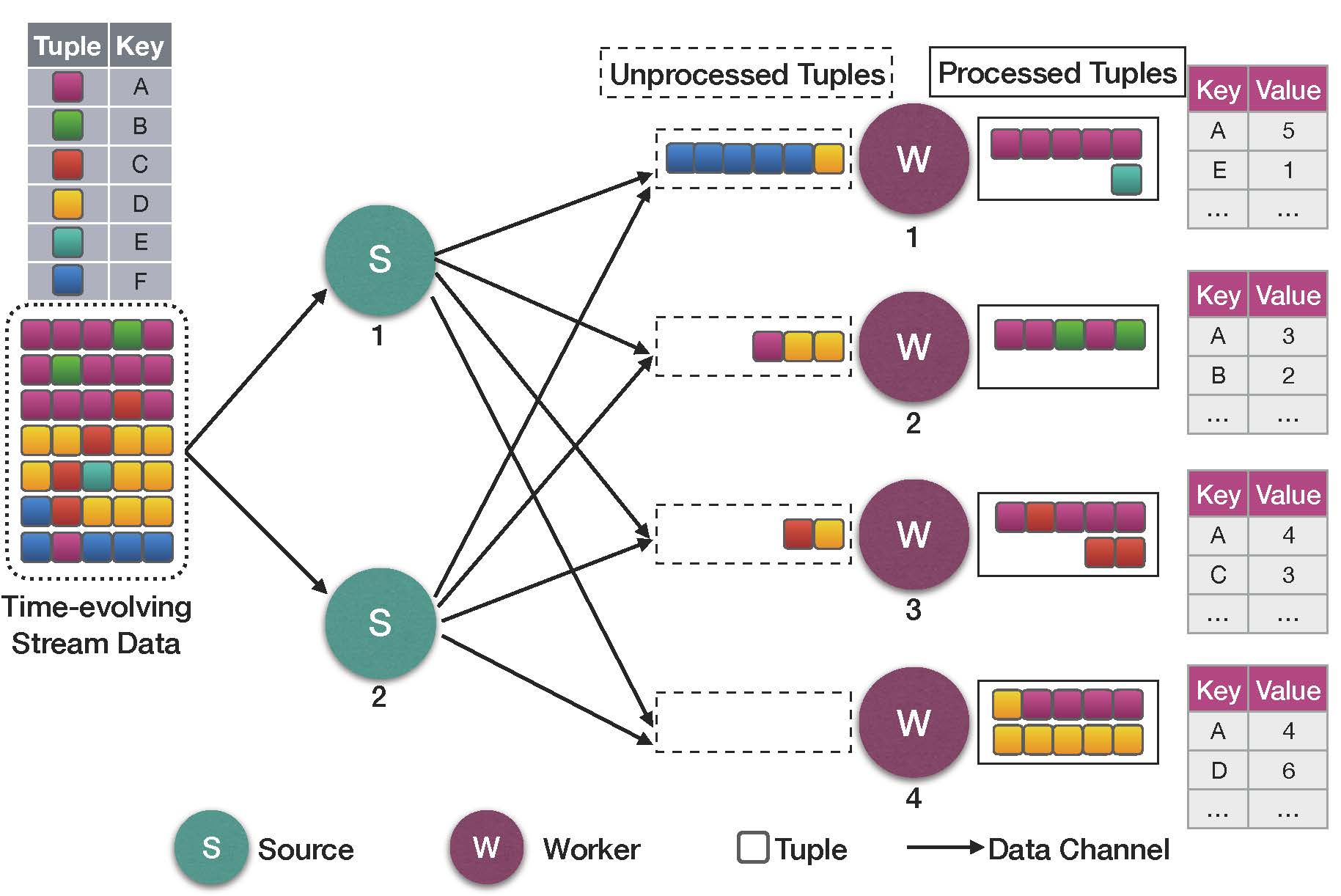}
  \caption{The typical workflow of distributed stream processing. Each colored tuple indicates a unique key. The tuples of time-evolving stream dataset start inflowing the sources from the upper-right corner.
  }
  \label{fig:DAG}
\end{figure}

Distributed stream processing engine (DSPE)~\cite{Storm,Flink,zaharia2013discretized,Samza,neumeyer2010s4} often runs on a cluster of machines that can communicate with each other via messages. The target stream applications are processed under these DSPEs in the form of a directed acyclic graph (DAG). Figure~\ref{fig:DAG} depicts a typical workflow of DSPE for the top-$k$ {\sf word count} stream application\footnote{Word count is a simple program that counts the number of occurrences of each word in a given input stream data} based on DAG where the vertex represents the operator of the stream engine that is applied on an incoming data stream for the data transformation. The directed edge represents data channel that points from an upstream operator (also called {\em source} for short) to a downstream operator (called {\em worker} for short). The data flow along these edges, representing a series of tuples, each associated with a key.

In order to achieve high performance, DSPE usually exploits data parallelism by running many instances of these operators. Each operator is responsible for handling a set of partitioned input sub-stream data, which relies on the creation of a particular grouping scheme (as will be discussed in Section 2.2).
In this case, a well-known problem for DSPE is load imbalance. For the example in  Figure~\ref{fig:DAG}, the key for each tuple is the word itself. Sources distribute tuples to workers based on a specific grouping scheme. Workers count the occurrence number of each word. The hot-key $F$ in this time-evolving stream data may be identified as non-hot potentially, leading to imbalanced tuple assignment. Also note that keys often have been duplicated in different workers with proportional memory overhead to the number of word types. The inefficiency of these aspects will be extensively investigated in Section 2.3.

\subsection{Existing Stream Grouping Schemes}

\indent The input stream is composed of a sequence of tuples, each of which is associated with a key. As shown in Figure~\ref{fig:DAG}, different colored tuples correspond to different keys. Grouping scheme assigns each tuple to a worker by key. Different grouping schemes may make different decisions for this key assignment, with a summary as follow:

\begin{itemize}[leftmargin=*]
    \item \bfseries Shuffle Grouping (SG)\mdseries~\cite{Storm}: This scheme sends each tuple from the source to a round-robin selected worker, ensuring that each worker can evenly have the tuples.
    \item \bfseries Field Grouping (FG)\mdseries~\cite{Storm}: This scheme ensures that the same key is always sent to the same worker via hashing.
    \item \bfseries Partial Key Grouping (PKG)\mdseries~\cite{nasir2015power}:  This scheme can be treated as a bounded FG. A given key for the PKG is allowed to be processed by two workers at most. 
    \item \bfseries D-Choices (D-C)\mdseries~\cite{nasir2016two}: This scheme is an improved PKG, which allows that frequent keys can be processed by $d$ workers at most where $d$ is determined by the distribution of key. Other keys continue using PKG.
    \item \bfseries W-Choices (W-C)\mdseries~\cite{nasir2016two}: This scheme is similar to D-C, and the only difference is that it allows frequent keys can be processed on the entire set of workers instead of $d$ ones.
\end{itemize}

\begin{figure*}[t]
  \centering
  \includegraphics[width=7.0in]{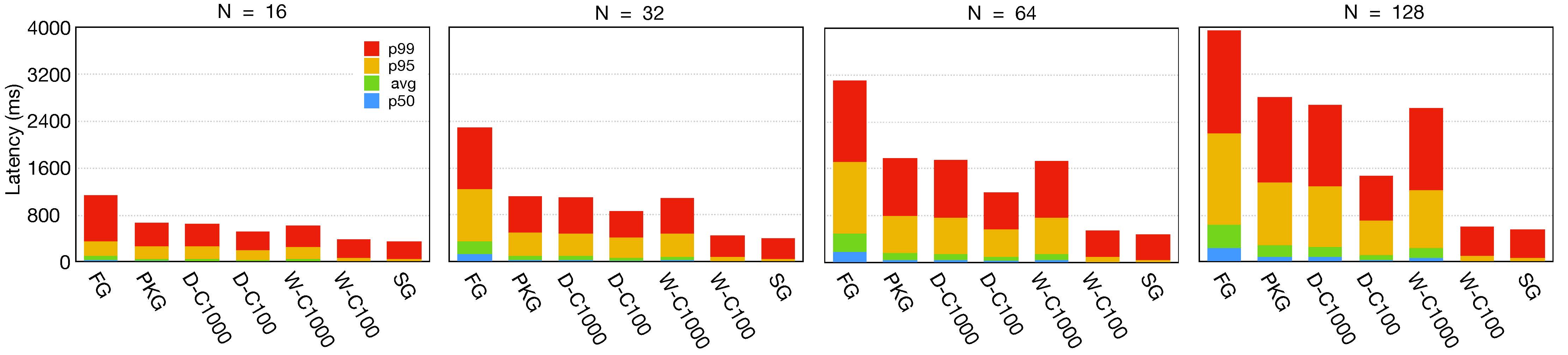}
  \vspace{-1em}
  \caption{Latency of FG, PKG, SG, D-C and W-C on the Amazon Movie Review dataset with different number of workers. D-C100 and D-C1000 indicate different maximum set of keys by 100 and 1000, respectively. W-C has the similar denotation.}
  \vspace{-1em}
  \label{fig:MO_LA}
\end{figure*}

\begin{figure}[t]
  \centering
  \includegraphics[width=3.33in]{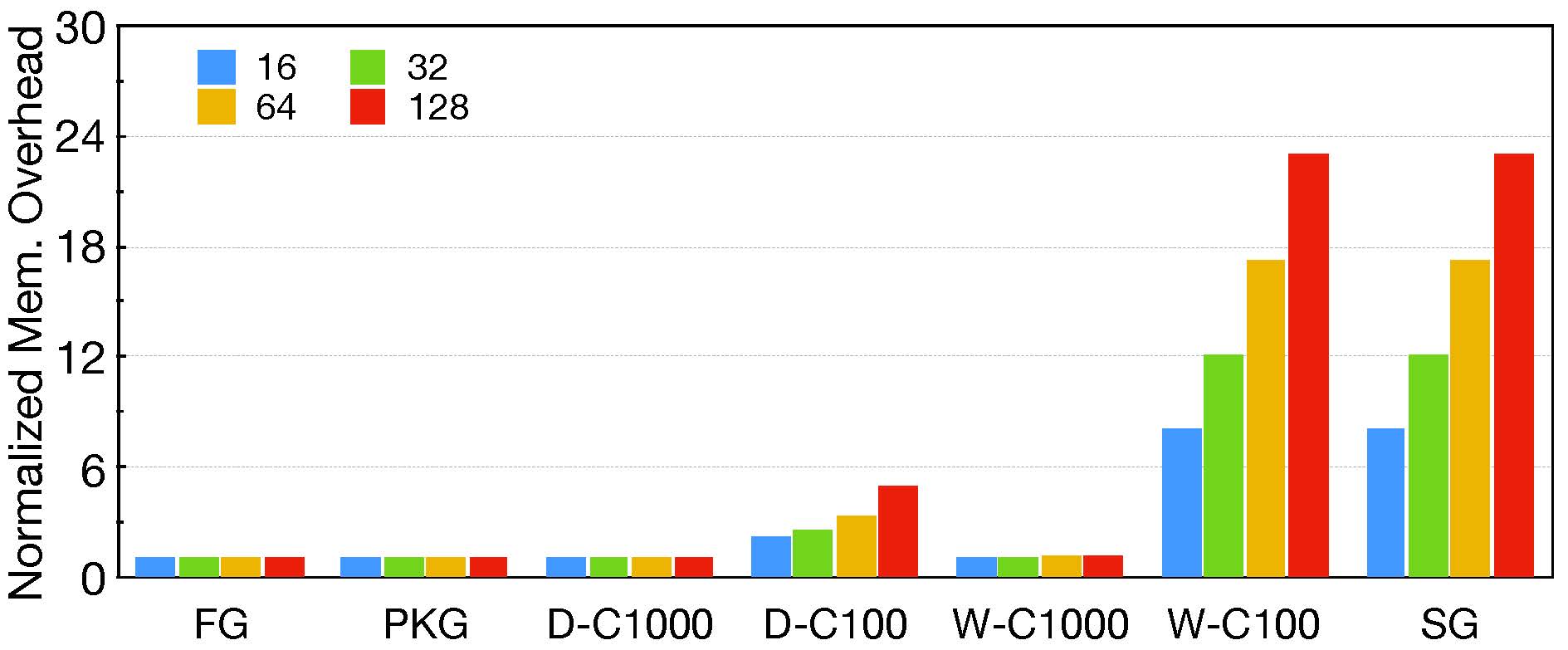}
  \caption{Normalized memory overhead of FG, PKG, SG, D-C and W-C with different number of workers. All results are normalized to FG.}
  \label{fig:MO_MEM}
\end{figure}

\subsection{Issues with Existing Grouping Schemes on Time-evolving Stream Datesets: A Motivating Study}\label{sec:Motivation Study}

These previous efforts~\cite{nasir2015power,nasir2016two} have made a significant advance for the load balance problem of DSPEs, particularly for skewed stream data. By considering the hotness of keys from the entire processing lifetime, their original key identification and assignment, however, are in essential unaware of the frequency variation of hot key within a bounded time interval. As a result, existing grouping schemes may result in the potential issues for the time-evolving stream processing with either load imbalance or prohibitive memory overhead, which can be much serious at scale (with a large number of workers).

To investigate this problem, we have conducted a set of experiments on the real-world time-evolving Amazon Movie Review stream dataset\footnote{Amazon movie review dataset collects the movie popularity, which can be significantly varying for different time periods.} with different machine scales (16, 32, 64 and 128 workers) for {\sf word count} application based on different grouping schemes discussed in Section 2.1. Note that we test D-C and W-C schemes by considering top-$100$ and top-1000 keys in this motivating study.

\textbf{Load Imbalance Issue}\quad
Figure~\ref{fig:MO_LA} depicts the results of latency, which is widely used for representing the load balance of the DSPEs~\cite{xing2005dynamic,nasir2015power,nasir2016two}. The lower the latency is, the more balanced the system is. The 99th percentile latency of FG and PKG is up to 3,945 and 2,808 milliseconds, respectively. Both FG and PKG have high latency because of assigning only one or two workers to each key. The skew distribution of the key results in extreme load imbalance of each worker. The latency of W-C and D-C is related to the number of statistical keys. If there are 1000 keys, latency of both W-C1000 and D-C1000 is almost the same as the PKG. With the increase of workers, the latency has a significant increase. This is due to inaccurate identification in the sense that some hot keys are detected as non-hot. If 100 keys are involved, the latency of D-C100 and W-C100 can have a part of improvement, but scalability issue below arises.


\textbf{Scalability Issue}\quad
Figure~\ref{fig:MO_MEM} depicts the results of memory overhead. FG assigns only one worker per key, and hence, it has little memory overhead as can be seen in Figure~\ref{fig:MO_MEM}. In contrast, we can see that SG has the highest memory overhead by up to 23.16x in the case of 128 workers since many states have been replicated. The D-C100 and W-C100 are similar to the SG. When the number of workers increases, the memory overhead increase significantly. This is due to inaccurate identification in the sense that some non-hot keys are detected as hot. Therefore, SG, D-C100, and W-C100 may suffer from scalability problems. To ensure system scalability, we set the maximum set of keys by $1000$ for the following experiment.

{\bf Summary} It can be seen that neither of existing grouping schemes can perform well in both load balance and scalability. Although state-of-the-art D-C and W-C schemes have made the advance for a relatively good tradeoff, they may be still far from the ideal situations (where SG scheme shows the optimal case for latency criteria while FG scheme represents the optimal case for memory overhead criteria). More importantly, their tradeoff gradually underperforms as the number of workers is increasing. There still lacks effective grouping scheme for efficiently processing these time-evolving stream data at scale.

\subsection{Challenges of Balancing Time-evolving Streaming Processing at Scale}\label{sec:Background and Motivation:challenge}
Time-evolving stream data has a significant feature with the significant frequency variation of keys within different time intervals. Not only with the {\em global} load balance for the final state during the entire lifetime, time-evolving stream processing but also needs to additionally consider the {\em local} real-time load balance within some time interval at every moment, arising several unique challenges.



First, by considering the time-evolving factor, the identification scope for the hot keys has been consequently changed from the entire processing to a large number of short time intervals. The problem of identifying recent hot keys within a time interval has been extensively studied in the Data Mining field, which can fall into two broad categories. Sliding-window based approach~\cite{golab2003identifying,chang2003finding,arasu2004approximate,wang2005tfp,deng2012new} uses window threshold for bounding recent key counting. To get the accurate results, they have to use a large window size at the cost of potentially prohibitive memory overhead. Time-aware based approach~\cite{mabroukeh2010taxonomy,shan2009frequent,lim2014fast} proposes that recent items have more weights so that a stale item is more likely to be pruned than a recent one. This approach uses a replacement strategy to reduce memory overhead, but each update for all items requires a time weight modification, leading to a large amount of computation.

Nevertheless, we should note that time-evolving stream processing often involves a large number of recent hot key identification operations, which can be easily more than millions for the real-world stream dataset. Technically, each of these operations is supposed to be efficient and lightweight so that the whole DSPE system can spread their superiority for load balance and scalability.
There still lacks an effective technique to accurately identify the recent hot keys while preserving the low overhead in both computation and memory consumption.

Second, after the hot key identification, we have to assign an appropriate worker for each identified recent hot key. As discussed previously, the traditional stream processing only considers the global load balance for the final states. Thus, they simplify the work assignment problem by evenly assigning all tuples to the given workers. Nevertheless, the reality is that the processing capability between workers is often different for many reasons, e.g., heterogeneous devices or network delays. As a consequence, it is likely for existing approaches to assign the keys for a busy worker in some time interval, leading to the local imbalance. An ideal method for work assignment is to select the optimal candidate worker according to the number of unprocessed tuples and processing capacity of workers.

Nevertheless, it is extremely difficult, if not impossible, to make efficient worker assignment. The unprocessed tuples information of workers is usually located in remote with respect to the source. Frequently requesting the queue states from workers may lead to a large amount of communication overhead between sources and workers. More serious is that this requested information may be quickly out of data since
the state of workers is often changing dramatically. There remain tremendously challenging for developing such an efficient worker assignment for time-evolving stream processing.


\section{Overview}\label{sec:Overview}
To cope with the aforementioned challenges, we design our grouping approach in accordance with the following interesting observations for time-evolving stream processing.

\underline{\bf\em Observation 1}: {\em The occurrence frequency between the recent hot keys and non-hot keys in the time-evolving stream data remains a large difference with a skewed distribution.}

One typical example accounting for the above observation is {\sf twitter}  dataset. Although its catchword may change from one to
the other over time, the occurrence frequency of these catchwords can
be still significantly higher than the non-hot ones (within a short interval).

This finding has two implications at least for the recent hot-key identification. First, in spite of the frequent variation of hot keys, a small fraction of these keys can still dominate most loads during the whole stream processing. This allows to continue using ``eighty-two'' golden rule by handling these few critical keys for the balance of most loads. Since only a few keys are saved in multiple workers, a large amount of memory overhead can be saved. Second, hot keys are subject to change over time. The potentially hot keys may be inaccurately identified as non-hot ones from a global perspective in prior work~\cite{nasir2016two}. Considering the skewed distribution of hot keys in a short interval, this implies that it is supposed to identify recent hot keys accurately in a locally-bounded (instead of global) manner.


\underline{\bf\em Observation 2}: {\em Considering the operation type of stream processing are usually simplex, the processing time for the same batch of tuples under the same given worker can be considered same with a negligible performance difference.}

\begin{figure}[t]
  \centering
  \includegraphics[width=3.5in]{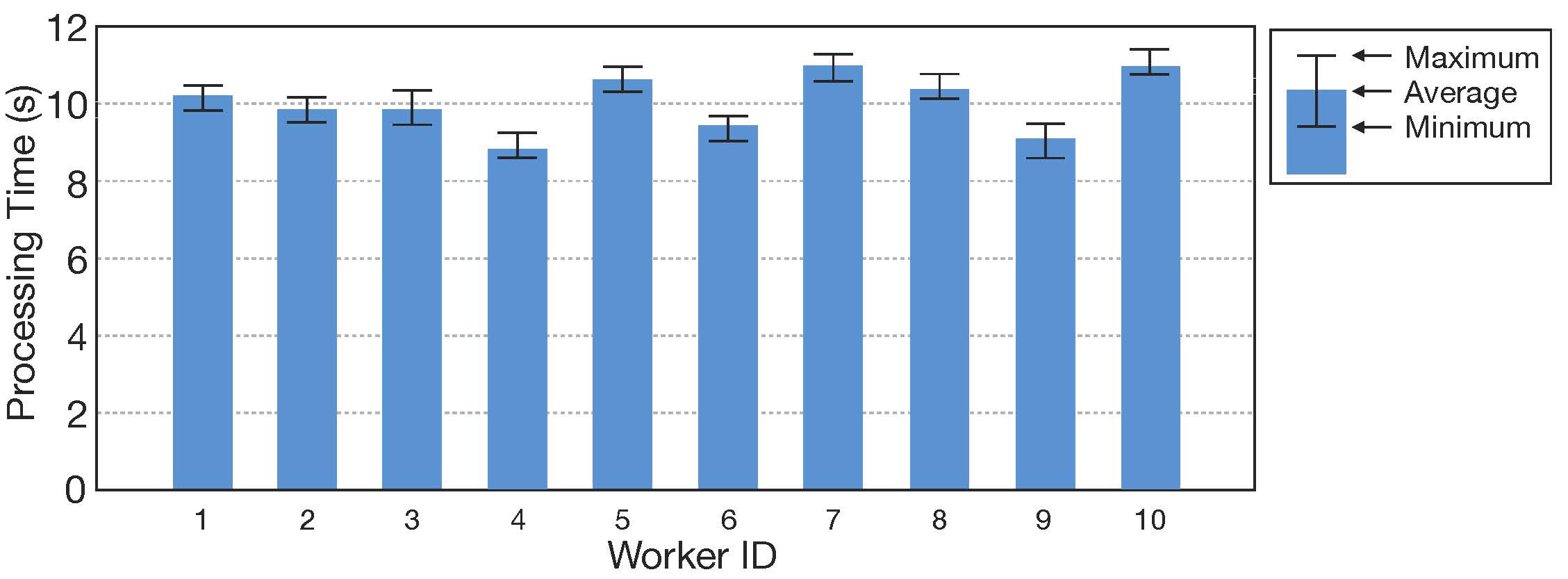}
  \vspace{-1.5em}
  \caption{The processing time for 10 workers. Each worker processes 50,000 tuples on the Amazon Movie datasets 12 times.}
  \label{fig:ob2}
  \vspace{-1.5em}
\end{figure}

\begin{figure}[t]
  \centering
  \includegraphics[width=3.5in]{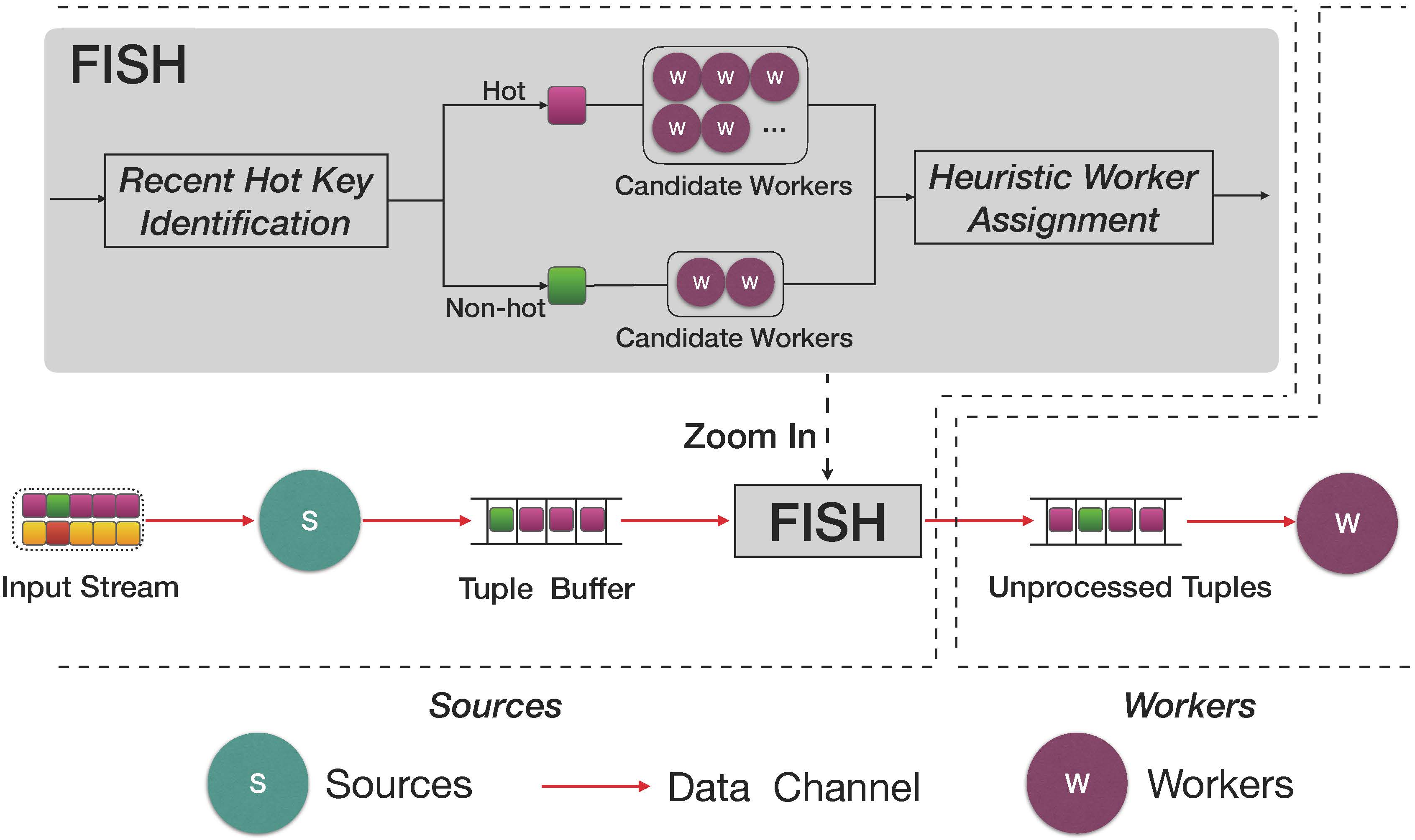}
   \vspace{-1.5em}
  \caption{FISH infrastructure relative to the streaming processing and its internal organizational structure ({\em our work is shaded})}
  \vspace{-1.5em}
  \label{fig:overview}
\end{figure}

Figure~\ref{fig:ob2} illustrates the performance results of processing every 50,000 tuples 12 times for 10 randomly-selected workers. We can see that the performance fluctuation range can be on average as small as 4.37\%, which can be often considered reasonable and negligible in practice~\cite{kapoor2004capprobe}.


This finding gives us an important implication for assigning an appropriate worker among all workers. The premise is that we have to know which worker has the fewest tasks unprocessed, which are generally unavailable at the source end. The intuitive method obtains this information via the considerable communication between workers. In contrast, this observation allows us to {\em infer} (rather than communicate) the unprocessed computation amount of all workers in a more efficient manner.


According to these implications, we propose a custom-made grouping approach with the specially-designed key identification and work assignment. Figure~\ref{fig:overview} illustrates the overview of our approach (named FISH), consisting of two major components as follows.

{\bf Accurate Recent Hot Key Identification} (Section 4.1): This part aims at accurately identifying the recent hot key for the time-evolving stream data. Although there exist a vast body of previous studies on recent hot key identification. These approaches are originally designed for mining the accurate data in data-mining applications, not yet satisfying the efficient requirement in the sense of low overhead in computation amount and memory consumption for stream processing applications.
We present a specialized recent hot key identification approach that can accurately identify hot keys for a recent time interval with low computational and memory overhead.

{\bf Heuristic Work Assignment} (Section 4.2): Given a set of workers, this part aims at assigning the identified hot keys to the appropriate workers for load balance. Unlike the previous studies that simply consider the global load balance at the final state (as discussed in Section 2.4), we additionally consider the local load balance at every time interval. This is particularly important for time-evolving stream processing. In contrast to communication-based worker assignment approach with heavy communication overhead, we propose a heuristic worker assignment, which can precisely infer the worker processing capacity based on the history information for worker assignment in a more efficient manner.

Note that this work focuses on addressing the common case where each tuple is associated with a single key. For the scenario where each tuple is allowed to carry multiple keys, we can still extend FISH to combine specific applications by prioritizing or synthesizing multiple keys. This is out of our scope, which can be interesting future work.

\begin{table}[t]
\begin{center}
\caption{Notations used in this work}
\label{sample-table}
\vskip 0.0in
\begin{tabular}{ll}
\hline
Symbol   &  Descriotion \\
\hline
\emph{$\alpha$}    &   time decaying factor \\
\emph{$\theta$}    &    threshold for the hot key identification\\
\emph{$c_{k}$} &  counter of a key ${k}$\\
\emph{$d_{min}$} &  minimal number of workers for hot key\\
\emph{$f_{top}$} &  the highest frequency\\
\emph{$f_{k}$} &  frequency of a key ${k}$\\
\emph{k,v}  &     key identifier\\
\emph{$t_{pri},t_{cur}$}  &     prior and current timestamp \\
\emph{w}  &     worker identifier\\
\emph{A}    &     set of candidate workers \\
\emph{$C_{w}$}    &     unprocessed tuples for worker $w$\\
\emph{D}    &   set of input stream data\\
\emph{K}    &   set of top frequent keys\\
\emph{$K_{max}$}    &   the maximum capacity of the set $K$\\
\emph{M}    &     set of assignable workers for each key\\
\emph{$N_{epoch}$}    &    the number of sequential tuples in an epoch \\
\emph{$N_{w}$}    &   the number of assigned tuples to worker $w$ \\
\emph{$P_{w}$}    &     the processing capacity for worker $w$\\
\emph{$T_{w}$}  & the estimate waiting time for worker $w$\\
\emph{$W_{num}$}  &  the number of workers\\
\hline
\end{tabular}
\end{center}
\end{table}

\section{Fish}\label{sec:Mechanism}
This section elaborates the design of the recent hot key identification and heuristic work assignment. For facilitating the descriptions, we define several notations used in this work. Table ~\ref{sample-table} lists the details regarding notations.

\subsection{Epoch-based Recent Hot-key Identification}\label{sec:Solution:Finding Recent Frequent Keys}
People often treat the hot key identification in the whole lifetime of stream processing. We either use a time-aware factor to compute the frequency of all keys~\cite{mabroukeh2010taxonomy,shan2009frequent,lim2014fast} with a large amount of computation, or use the additional storage to memorize the history frequency of all keys in the cost of memory overhead~\cite{golab2003identifying,chang2003finding,arasu2004approximate,wang2005tfp,deng2012new}.

Motivated by observation 1, the core idea of our recent hot key identification is an epoch-based approach, which divides the entire lifetime of stream processing into many epochs. {\em Epoch}  is a collection of sequential tuples. The intra-epoch counts the occurrence number of the key, which only stores the number of most frequent keys~\cite{manku2002approximate,karger2004simple} for reducing the prohibitive memory overhead. The inter-epoch frequency counting of keys uses a time-aware~\cite{mabroukeh2010taxonomy,shan2009frequent,lim2014fast} approach which adopts epoch-level (rather than tuple-level) update for reducing the superfluous amount of computation. Based on the frequency results, these keys are finally classified into hot and non-hot ones.


\begin{figure}[t]
  \centering
  \includegraphics[width=3.5in]{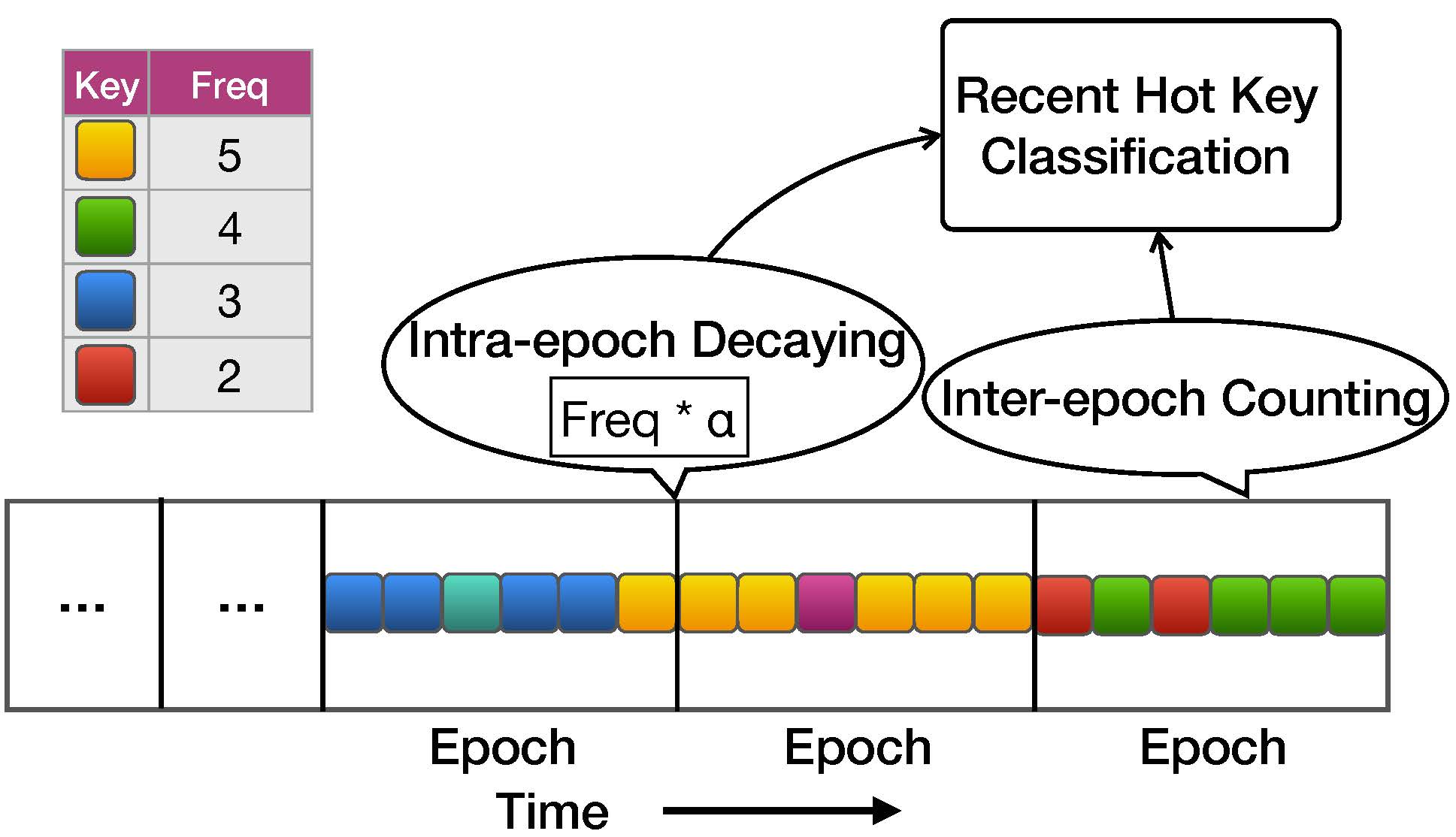}
  \vspace{-2em}
  \caption{Procedure of epoch-level recent hot-key identification}
  \vspace{-0.5em}
  \label{fig:epoch}
\end{figure}

\begin{algorithm}[t]
\caption{Epoch-based Key Frequency Statistics}
\DontPrintSemicolon
\KwIn{$\alpha$ -- time decaying factor\\
\hspace{9.4mm} $D$ -- input stream data\\
\hspace{9.2mm} $N_{epoch}$ -- the size of epoch\\
\hspace{9.2mm} $K_{max}$ -- maximum capacity of the set K}
 $K \leftarrow \phi$\;
 $counter \leftarrow 0$\;
\ForEach {k $\in$ D}{
    \tcc{Inter-epoch decaying}

    \If{$counter = N_{epoch}$}{
        TimeDecayingUpdate$(K)$\;
        $counter \leftarrow 0$\;
    }
    \tcc{Intra-epoch counter}
    \eIf{$k \in K$}{
        $c_{k} \leftarrow c_{k} +1$\;
    }{
        \tcc{Insert or replace the key}
        \eIf{$\left | K \right |  <  K_{max} $}{
        $K \leftarrow K\cup \left \{ k \right \}$\;
        $c_{k} \leftarrow 1$\;
        }{
        ReplaceMin$(K,k)$\;
        }
    }
    $counter \leftarrow counter+1$\;
}
\textbf{Subroutine} {ReplaceMin$(K,k)$}\;
\quad$k_{min}$ $\leftarrow$ $min_{v \in K}c_{v}$\;
\quad$K \leftarrow K\setminus \left \{ k_{min} \right \}\cup \left \{ k \right \}$\;
\quad $c_{k} \leftarrow c_{k_{min}}+1$\;
\textbf{Subroutine} {TimeDecayingUpdate$(K)$}\;
\tcc{Update counters according to the $\alpha$}
\ForEach {$v\in K$ }{
    $c_{v} \leftarrow c_{v} \times \alpha $\;
}
\end{algorithm}

\subsubsection{Key Frequency Statistics}
In the following, we next introduce how we obtain the frequency of keys based on an epoch-driven approach.

\textbf{Intra-epoch Frequency Counting}\quad The intra-epoch counting aims to count the occurrence number of the key in each individual epoch. To reduce memory overhead, we continue to only store the most frequent $K_{max}$ keys~\cite{karger2004simple}. The related descriptions are located between Lines 8-17 in Algorithm 1. When a new key appears, if the current number of keys stored in $K$ is less than the maximum capacity, this key will be merged into the $K$ set, and its occurrence number is incremented. If $K$ is full, we use a replacement strategy to replace the least counted key from $K$. Note that its occurrence number is set to that of replaced ones plus 1 rather than 1 (as shown in \textsf{ReplaceMin}). The major reason is just for avoiding the unreasonable replacement of new keys~\cite{karger2004simple}. To be more specific, if it is set to 1, once a new key comes, we will always replace this key until the occurrence number of this key exceeds others. This is unreasonable since the previous key is replaced and its valuable information is not reusable for the memory saving.


\textbf{Inter-epoch Hotness Decaying}\quad Instead of performing a time decaying update when each tuple arrives (as described between Line 5-7 in Algorithm 1), we adopt a time-aware decaying approach in the epoch granularity. After tuple statistics in each epoch is completed, we multiply the counters of all the stored keys by $\alpha$ ($0<\alpha<1$) so that the time decaying effect can be taken. Hence, the counter is not only related to the number of occurrence number of the key but also the time decaying factor.

It is worth noting that the size of the epoch directly determines the computational overhead of the recent hot-key identification. The larger the epoch size is, the lower the computational overhead is, and vice versa. Nevertheless, large epoch size may also affect the accuracy of the hot key identification. We conduct our experiments with the empirical epoch size of $1000$ by default. It is revealed that this result can cover almost all datasets (as will be discussed in Section 6) without compromising identification accuracy, and also can reduce the computational complexity of decaying updates by three orders of magnitude.


\begin{algorithm}[t]
\caption{Classification of Hot Key (CHK)}
\DontPrintSemicolon
\KwIn{$d_{min}$ -- minimal number of workers for hot key\\
\hspace{9.3mm} $f_{top}$ -- the highest frequency\\
\hspace{9.3mm} $f_{k}$ -- the frequency of the key $k$}
\KwOut{$d$ -- number of candidate workers}
\eIf{$f_{k} > \theta$}{
    \tcc{Assign the number of candidate workers to the key}
    $index \leftarrow  \lfloor\log_{2} / (f_{top} \ / f_{k}) \rfloor$\;
    $ d \leftarrow   W_{num}  / 2^{index} $\;
    \If {$d <  d_{min}$}{
        $d \leftarrow d_{min}$\;
    }
    \eIf {$M_{k} <  d$}{
        $M_{k} \leftarrow  d$\;
    }{
        $d \leftarrow M_{k}$\;
    }
}{
    $d \leftarrow 2$\;
}
\Return $d$.\;
\end{algorithm}

\subsubsection{Hot Key Classification}\label{sec:Solution:Classify Hot Keys}
We next introduce to classify recent hot keys based on the frequency results.
Algorithm 2 describes the procedure of hot key classification (denoted as CHK). To determine the number of workers to which each key can be assigned, we use the set $M$ to hold the number of the candidate workers for each hot key. We are based on the idea that the higher the frequency is, the more workers assigned. First, we get the number of arithmetic assignment workers for the hot key through the formula from line 1 to 4 in Algorithm 2. Second, if the value of $d$ obtained is less than the minimum value of $d_{min}$, we directly assign $d$ to $d_{min}$. The $d_{min}$ is related to the sum of the frequency of all hot keys. Then, considering that the frequency of the key changes, $M_{k}$ saves the number of workers previously assigned to key $k$. If $d$ is greater than the $M_{k}$, $M_{k}$ is updated to $d$ and $d$ workers are assigned to the key. Otherwise, we assign $M_{k}$ workers to the hot key. It is worth noting that we assign workers for each key through a consistent hash so that we can deal with the dynamic workers. The detailed contents will be introduced in Section 5.

\begin{algorithm}[t]
\caption{Heuristic Worker Assignment}
\DontPrintSemicolon
\KwIn{$A$ -- set of candidate worker\\
\hspace{9.4mm} $T$ -- time interval
}
\KwOut{$appro$ -- number of selected worker.}
$appro \leftarrow -1$\;
\tcc{Estimate the current status of each worker}
$t_{cur} \leftarrow GetNowTime( )$ \;

\If{$t_{cur} - t_{pri} > T$}{
    \ForEach{$w \in W$}{
        \eIf{$(C_{w}+N_{w}) \times P_{w} > T$}{
            $C_{w} \leftarrow ((C_{w}+N_{w}) \times P_{w} - T) / P_{w}$\;
        }{
            $C_{w} \leftarrow 0$\;
        }
    }
    $t_{pri} \leftarrow t_{cur}$\;
}

\tcc{Select the appropriate load worker}

\ForEach{$w \in A$}{
    \eIf{$appro\ = -1$}{
        $appro \leftarrow w$\;
    }{
        \If{$C_{appro} \times P_{appro} > C_{w} \times P_{w}$}{
            $appro \leftarrow w$\;
        }
    }
}
$C_{appro} \leftarrow C_{appro} + 1$\;

\Return $appro$\;

\end{algorithm}

\subsection{Heuristic Worker Assignment}\label{sec:Solution:Choosing a Light Load Worker}
This section introduces how to assign the identified hot keys to $d$ or $2$ workers by CHK. Choosing a light-load worker from $d$ or $2$ candidate workers is the next question that has to address. We present a heuristic method to efficiently estimate (rather than communicate in prior efforts) the runtime states of workers in a fine-grained time interval.

\subsubsection{Worker State Estimation}

\indent In order to fully mobilize each worker, each tuple is expected to be processed as soon as possible. The selection of the light load worker usually depends on two states of the worker: the number of unprocessed tuples and processing capacity. Unfortunately, obtaining this information from all workers can cause prohibitively communication overhead. We observe that stream processing usually takes the same kind of operation for processing each tuple. Therefore, we obtain the processing capacity ({\em the average processing time of a tuple}) of workers by a periodic sampling. Since the number of tuples for each worker can be directly obtained at the source end, we estimate that the number of unprocessed tuples of workers is as follow:$$C_{w} = \left ( (C_{w} + N_{w})\times P_{w} - T \right )/ P_{w} \eqno{(1)}$$
where $N_{w}$ is the number of assigned tuples from sources. $P_{w}$ is the processing capacity of the worker $w$ and $T$ is the fixed time interval (10s). As shown in Figure~\ref{fig:ob2}, there is little difference in the processing time for the same batch of tuples under the same worker. We set the default time interval to 10 seconds. We thus can estimate the number of unprocessed tuples $C_{w}$ for the worker $w$.


\subsubsection{Candidate Worker Selection}
We estimate the number of unprocessed tuples in a heuristic fashion. Each tuple is expected to be processed as quickly as possible to fully squeeze each worker for load balance. Considering potentially-different processing capability of different workers, 
we select the worker with the shortest waiting time as shown between Line 12 to 18 in Algorithm 3. The estimated waiting time can be expressed as follow:$$T_{w} =C_{w} \times P_{w} \eqno{(2)}$$
where $T_{w}$ is to estimate the waiting time for the worker $w$. Considering the similarity of stream processing, capture the states of workers using a sampling technique~\cite{warwick1975sample}.

\subsubsection{Example Illustration}
\begin{figure}[t]
  \centering
  \includegraphics[width=3.5in]{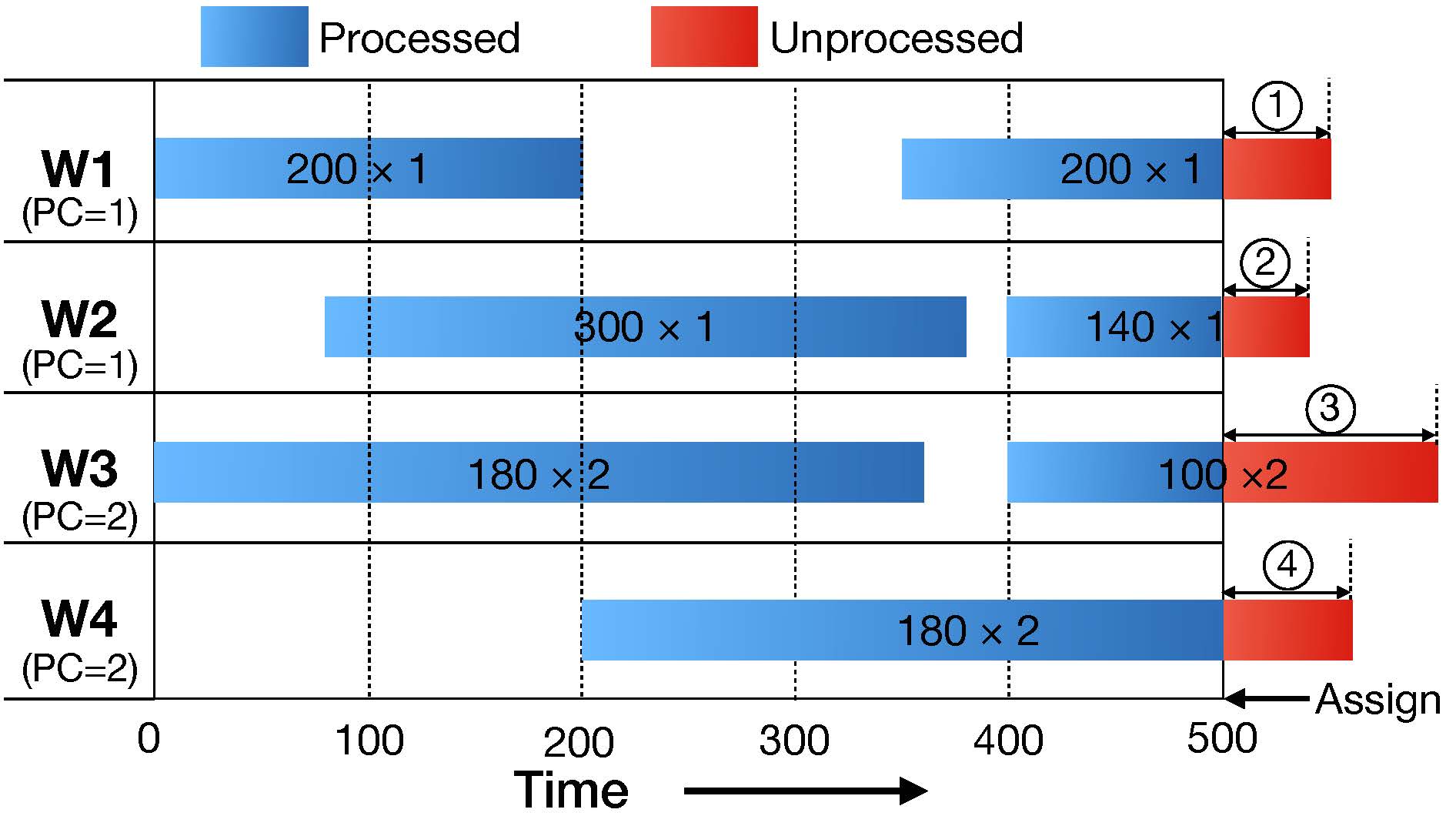}
  \vspace{-2em}
  \caption{Example of worker assignment. The bar indicates the processing status of tasks on different workers as the execution time goes by. Each bar is associated with $a \times b$ where $a$ denotes the number of tuples and $b$ indicates the processing capability (PC) of the worker.}
  \label{fig:assign}
\end{figure}

Figure~\ref{fig:assign} shows the example of how an appropriate worker is assigned. In this example, there are a total of 4 workers. Suppose the processing capacity of all workers are normalized to workers $W1$ and $W2$. Workers $W3$ and $W4$ have the twice processing capacity (PC) than $W1$ or $W2$.

Suppose the current time is at 500. What we need to do is to assign a tuple to a worker from $W1$, $W2$, $W3$, and $W4$. In Figure~\ref{fig:assign}, the blue bar indicates the time spent in processing tuples. The red bar represents the time required for the unprocessed tuples. The $W1$, $W2$, $W3$, and $W4$ are assigned 400, 440, 280, and 180 tuples, respectively.
Simply based on the number of assigned tuples as done in previous studies~\cite{nasir2015power,nasir2016two}, the worker $W4$ will be selected. In contrast, our work considers both unprocessed tuples and processing capacity for each worker. It is estimated that the waiting times for $W1$, $W2$, $W3$, and $W4$ are 50 (\circled{1}), 40
(\circled{2}), 100
(\circled{3}), and 60
(\circled{4}), respectively. We hence select $W2$ because of its shortest pending time, which is preferable over $W2$ for the subsequent tuple processing.

\begin{figure}[t]
  \centering
  \includegraphics[width=3.0in]{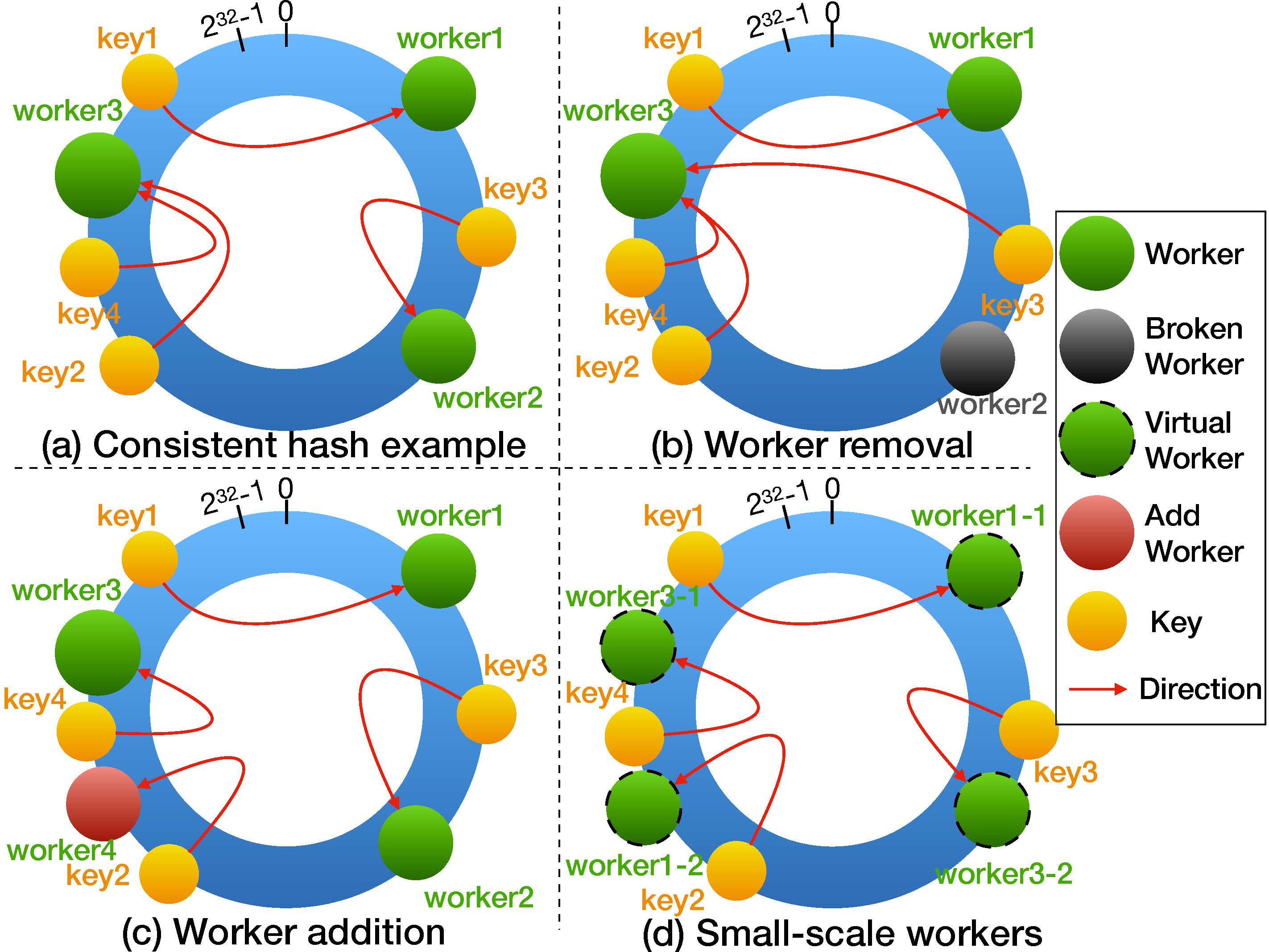}
  \vspace{-1em}
  \caption{An example of maintaining hashing consistency. {\bf(a)} A consistent hashing example with 3 workers; {\bf(b)} A case of removing a worker; {\bf(c)} A case of adding a worker; {\bf(d)} Small-scale worker deployment.}
  \vspace{-1.5em}
  \label{fig:CH}
\end{figure}

\section{Extension: Dynamic Change of Workers}\label{sec:Extension}
There remains the fact that the number of workers may be dynamically changing in a practical deployment. For example, a worker might be shut down or failed. Alternatively, the new worker is put into operation.

A typical approach for adapting the dynamic scenario is to use a hashing algorithm~\cite{eastlake2001us}. By using a hash function $F =$ HASH $(k)$ mod $n$ where $k$ is the key and $n$ is the number of workers, keys can be mapped to different workers. Nevertheless, the overhead of this simple mapping is subject to the number of workers. When a worker is removed or added, all keys have to be remapped to all workers, resulting in considerable memory overhead.

An alternative approach is that we can create a virtual ID mapping table for the workers based on a maximum number of supported workers, and make the assignments based on virtual IDs~\cite{plaxton1996fast}. However, this approach suffers from two defects at least. First, modifications to the virtual ID mapping table may introduce a large amount of synchronization overhead for the consistency across all sub-streams. Second, the assignment of workers is not random, resulting in the key not being evenly distributed to workers. It directly affects the balanced distribution of the load.


Let us reconsider this problem, which can be abstracted to map a batch of keys to $n$ workers and need to meet two requirements. First, all keys are supposed to be randomly and mapped evenly to workers. Second, the addition or reduction of workers does not cause a large number of key-to-worker re-mappings with  monotonicity.
We therefore propose to use consistent hash~\cite{karger1997consistent,karger1999web} for reducing the unnecessary key-to-worker mappings.

Figure~\ref{fig:CH} shows a case of consistent hashing algorithm. Each key can be hashed to a space with $2^{32}$ buckets\footnote{The size of the bucket space is determined by the hash algorithm. The hashing algorithm~\cite{eastlake2001us} is used in our method to return 32-bit integer data. The maximum value of unsigned integer data is $2^{32}-1$, we thus use $2^{32}$ for bucket space.}. We connect these numbers to form a hash ring. The data is mapped to the ring through the hash algorithm. Now we hash \emph{key1}, \emph{key2}, \emph{key3}, and \emph{key4} to the hash ring through a specific hash function. The worker is also mapped to the ring by using the same hash algorithm as key. In a clockwise direction, all the keys stored in their nearest worker. In Figure~\ref{fig:CH}(a), the current state should be that \emph{key1} is stored in \emph{worker1}, \emph{key3} in \emph{worker2}, \emph{key2} and \emph{key4} in \emph{worker3}.

\textbf{Worker Removal and Addition} Suppose the \emph{worker2} is crashed. As shown in Figure~\ref{fig:CH}(b), we have to remove it from the hash ring. According to the clockwise rule, \emph{key3} is then mapped to \emph{worker3}. No changes for all other keys have happened. Alternatively, suppose a new worker is added.  Figure~\ref{fig:CH}(c) illustrates the way for this case where \emph{worker4} is added. By the clockwise shift rule, \emph{key2} is originally mapped to \emph{worker3} and will now be remapped to \emph{worker4} as \emph{worker4} is closer to \emph{key2} than \emph{worker3} on the ring. The other key still maintains the original mapping relationship with only change for the \emph{key2} mapping. In summary, the addition or removal of workers only affects the mapping of keys with a few steps (by just only changing worker to adjacent worker) on the hash ring. Correspondingly, a small portion of the keys in the ring space need to be remapped.

{\bf Small-scale Worker Deployment}\quad Note that in the case that the number of workers is small, consistent hashing algorithm prone to causing the uneven distribution of keys for each worker. As shown in Figure~\ref{fig:CH}(b), when \emph{worker2} is removed with only two workers, \emph{key2}, \emph{key3}, and \emph{key4} will be mapped to \emph{worker3}. Only \emph{key1} will be mapped to \emph{worker1}. We complement to use a virtual node mechanism~\cite{karger1999web,stoica2003chord}, which calculates multiple hash values for each worker. By this means, each worker has multiple  virtual nodes, which are further mapped onto the hash ring. Figure~\ref{fig:CH}(d) shows an example with two virtual nodes for each worker. There thus have four virtual nodes, denoted as \emph{worker1-1}, \emph{worker1-2}, \emph{worker3-1}, and \emph{worker3-2}, respectively. The new key-to-worker mapping relationship in Figure~\ref{fig:CH}(d) (i.e., \emph{key1} and \emph{key2} are mapped to \emph{worker1}; \emph{key3} and \emph{key4} are mapped to \emph{worker3}) demonstrates that the distribution of keys is more balanced than otherwise.

\section{Evaluation}\label{sec:Evaluation}
\indent In this section, we evaluate the efficiency and effectiveness of FISH by answering five research questions:

\begin{table}[t]
\begin{center}
\caption{Time-evolving stream datasets}
\label{dataset-table}
\vspace{-1em}
\begin{tabular}{lllll}
\hline
Dataset   &  Abbr. &  Tuples & Keys\\
\hline
MemTracker    &  MT   &   49.21M     &   0.39M \\
Amazon Movie &  AM   &    7.91M   &   0.25M \\
\hline
Zipf     &   ZF  &    50M   &   $10^{5}$ \\
\hline
\end{tabular}
\vspace{-2em}
\end{center}
\end{table}

\begin{itemize}[leftmargin=*]
 \item {\bf\em RQ1}: How efficient is FISH compared to existing state-of-the-art grouping schemes? (Section~6.2)
  \item {\bf\em RQ2}: How to decide the internal parameters of FISH for load balance? (Section~6.3)
 \item {\bf\em RQ3}: How effective is each part of FISH? (Section~6.4)
 \item {\bf\em RQ4}: How effective is consistent hashing algorithm for dynamic extension of worker variation? (Section~6.5)
 \item {\bf\em RQ5}: How is overall effect of FISH for a practical deployment on Apache Storm? (Section~6.6)
\end{itemize}

\subsection{Experimental Setup}
\textbf{Simulation Settings}\quad We process the stream dataset by simulating the basic DAG in Figure~\ref{fig:DAG}. Sources extract the data and the workers perform the data aggregation. The input stream data is received by sources through shuffle grouping. Each data consists of a timestamp and a corresponding key. We assign each tuple to the specified worker based on the grouping scheme we desire to evaluate.

\textbf{Datasets}\quad We evaluate FISH on both real-world and synthetic stream datasets, as shown in Table~\ref{dataset-table}. We use two real-world datasets, including MemeTracker (MT)~\cite{leskovec2009meme} and Amazon Movie Review (AM)~\cite{mcauley2013amateurs}. MT provides quotes and phrases from blogs and news media. We consider a keyword stream, which consists of words in the quotes and phrases where 571 stopwords provided in~\cite{lewis2004rcv1} are excluded. AM provides user reviews with product identification, which is used as the key for the tuples.

As for synthetic Zipf (ZF) dataset, we generate 50M tuples with $10^5$ unique keys. Considering the skewness of stream data, the generated time-evolving ZF dataset has the following distribution with the exponent in the range $z\in$ $\left \{ 1.0,1.1,\dots,2.0 \right \}$. 1) For the first 0.8$\times N$ tuples, the occurrence probability of a given key $i$ obeys $Pr\left [i \right ]\propto i^{-z}$; 2) For the last ($1-0.8$)$\times N$ tuples, the occurrence probability of a given key $i$ obeys $Pr\left [i \right ]\propto (k-i+1)^{-z}$ where $k$ is $10^4$ and $N$ is 5M. To simulate the feature of time-evolving data, the algorithms have been run 10 times with a different seed for the pseudo-random number generator.

\indent {\bf Measurement Metrics}:
To evaluate the scalability, we use the amount of memory overhead all workers have totally consumed as the metric. The less total memory overhead of all workers incurs, the fewer memory duplicates have been caused, indicating the better scalability.

We use the processing time of loads to evaluate the load balance in the simulation environment. Generally, the more balanced the loads are, the better the worker can be fully mobilized. The execution time of different grouping schemes basically depends on the utilization of the workers, which can be used as an effective metric to represent the effect of load balance.

\begin{figure}[t]
  \centering
  \includegraphics[width=3.3in]{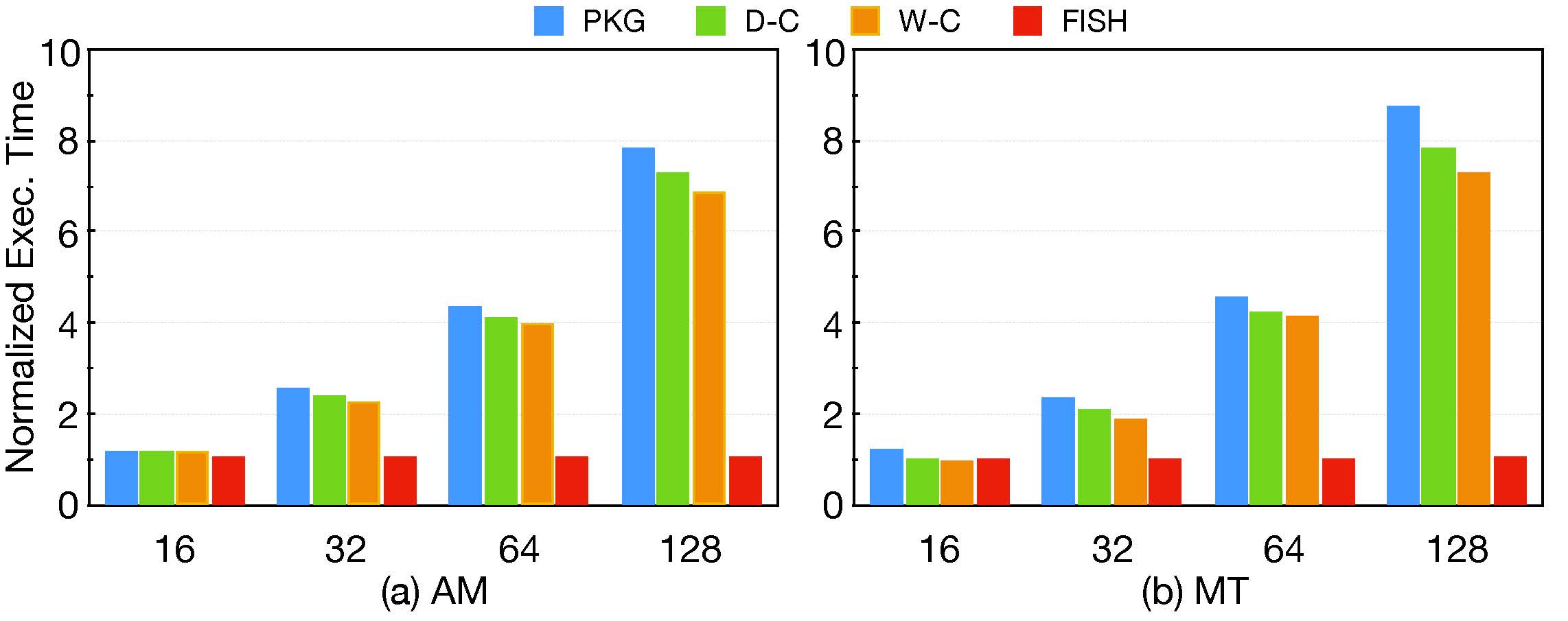}
\vspace{-1em}
  \caption{Execution time of PKG, D-C, W-C, and FISH with different number of workers on the real-world datasets. (a) is for AM, and (b) is for MT. All results are normalized to SG.}
\vspace{-1em}
  \label{fig:PR}
\end{figure}

\begin{figure*}[t]
  \centering
  \includegraphics[width=7.0in]{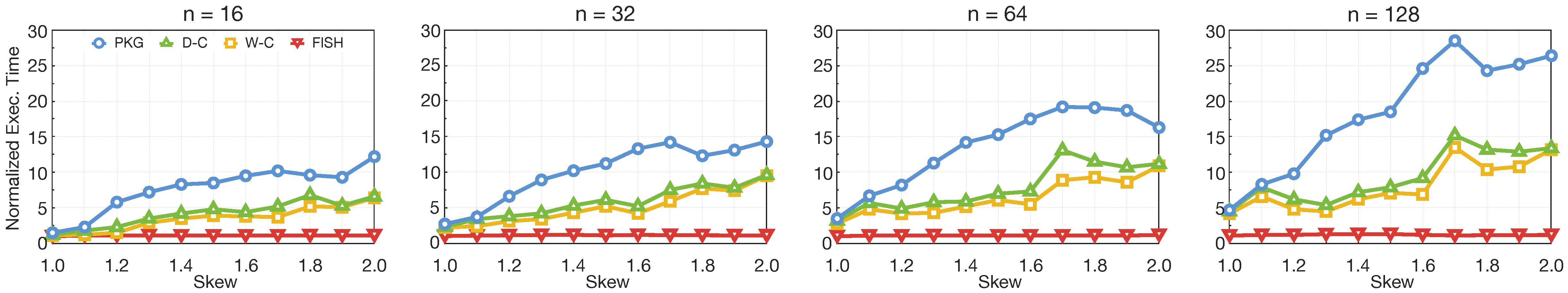}
  \vspace{-1em}
  \caption{Execution time of PKG, D-C, W-C, and FISH with different number of workers on the ZF datasets. All results are normalized to SG.}
  \vspace{-1em}
  \label{fig:PZ}
\end{figure*}

\begin{figure*}[t]
  \centering
  \includegraphics[width=7.0in]{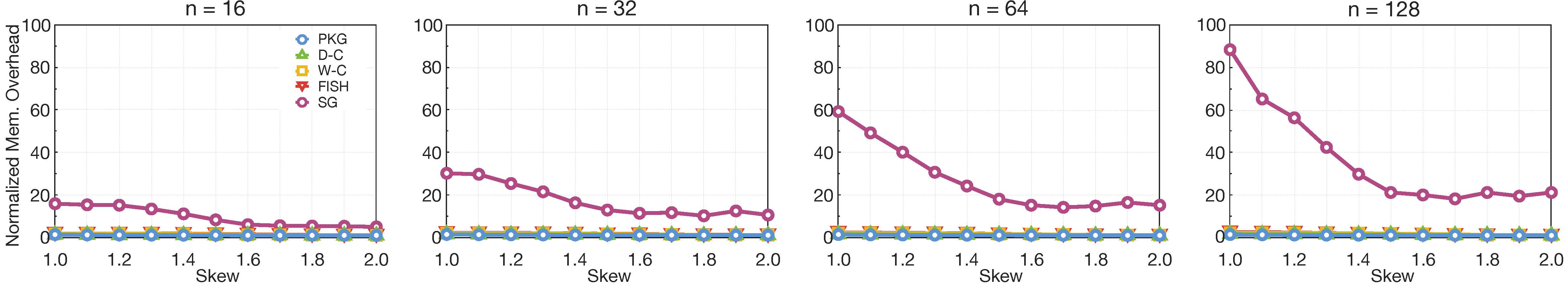}
\vspace{-1em}
  \caption{Memory overhead of PKG, D-C, W-C, and FISH with different number of workers on the ZF datasets. All results are normalized to FG.}
    \vspace{-1em}
  \label{fig:MZ}
\end{figure*}

\subsection{RQ1: Overall Evaluation}
We investigate overall load balance and memory overhead of FISH against state-of-the-art PKG, D-C, and W-C grouping schemes on both synthetic and real-world datasets. For the load balance, we use the SG as the baseline, which is a well-known grouping scheme with an ideal load balancing effect. For memory overhead, we use FG as a baseline since it does not generate any extra memory overhead.


\textbf{Load Imbalance}\quad Figure~\ref{fig:PR} illustrates the results on the real-world {AM} and {MT} datasets. We use the SG as the baseline. The lower the execution time is, the better the load balancing effect is. Compared to four tested grouping schemes, we can see that FISH has the best load balance effect for both {MT} and {AM} datasets. The execution time of FISH is almost same as the SG with the worst case of 1.07x. Compared to PKG, as the number of workers increases, the effect of FISH increases more significantly (from 1.19x to 8.32x for MT and from 1.12x to 7.31X for AM). This is because that PKG only assign two workers for all keys. The skew distribution of keys causes the tuples to be unevenly distributed among workers, resulting in load imbalance. Although W-C and D-C take into account the skew distribution of keys, its effect is still limited as the number of workers increases. Overall, FISH has up 7.44x and 6.95x improvement than D-C and W-C respectively. This is due to the fact that the feature of the time-evolving of the stream data is not taken into consideration, resulting in inaccurate hot key identification and inappropriate assignment.


Figure~\ref{fig:PZ} further investigates the load balance of FISH on synthetic ZF dataset with the different skew factor. Overall, the gap between the four grouping schemes is increasing with the number of workers increases.
PKG is worst among all of four grouping schemes. The effect of PKG becomes worse with the skew increases because it only assigns two workers for each key without considering the case of skewed data. The execution time of D-C and W-C becomes longer with the skew increases, although the skewed data feature is considered. Particularly with the increasing number of workers, the effect would become worse. FISH is up to 13.57x and 12.05x improvement than D-C and W-C respectively. This is because that time-evolving feature is not considered in D-C and W-C. As a result, they may lead to the fact that hot keys cannot be accurately identified and assigned. We also note that as the number of workers is scaling, FISH can always have the comparable load balance effect to SG with the worst case of 1.32x.

\textbf{Memory Overhead}\quad Figure~\ref{fig:MZ} shows the memory overhead of FISH compared to existing grouping scheme FG, SG, PKG, D-C, and W-C. For system scalability, not only load balancing but also memory overhead must be taken into consideration. We use the memory overhead of FG as a balance to normalize the results of other grouping schemes. FG assigns only one worker per key without any extra memory overhead. Thanks to the special assignment for a small fraction of keys, which dominate most loads in stream data. Even with the extended number of workers, the memory overhead of FISH is comparable (from 1.11x to 2.61x) to FG with 128 workers. Although SG is able to balance the load well with the increasing number of workers, the memory overhead has increased significantly (from 15.52x to 88.32x). Compared to FG, the memory overhead of PKG, D-C, and W-C schemes is close. Yet, they suffer from the problem of load imbalance, as depicted in Figure~\ref{fig:PR} and Figure~\ref{fig:PZ}. In summary, compared to all of existing grouping schemes, FISH showcases the best results in load balance and memory overhead for time-evolving stream data.

\subsection{RQ2: Internal Parameter Decision}
We next investigate how to decide the appropriate internal parameters of FISH for better effect. Two major parameters include the decaying factor $\alpha$ in Algorithm 1 and the hot key threshold $\theta$ in Algorithm 2. 

\begin{figure*}[t]
  \centering
  \includegraphics[width=7.0in]{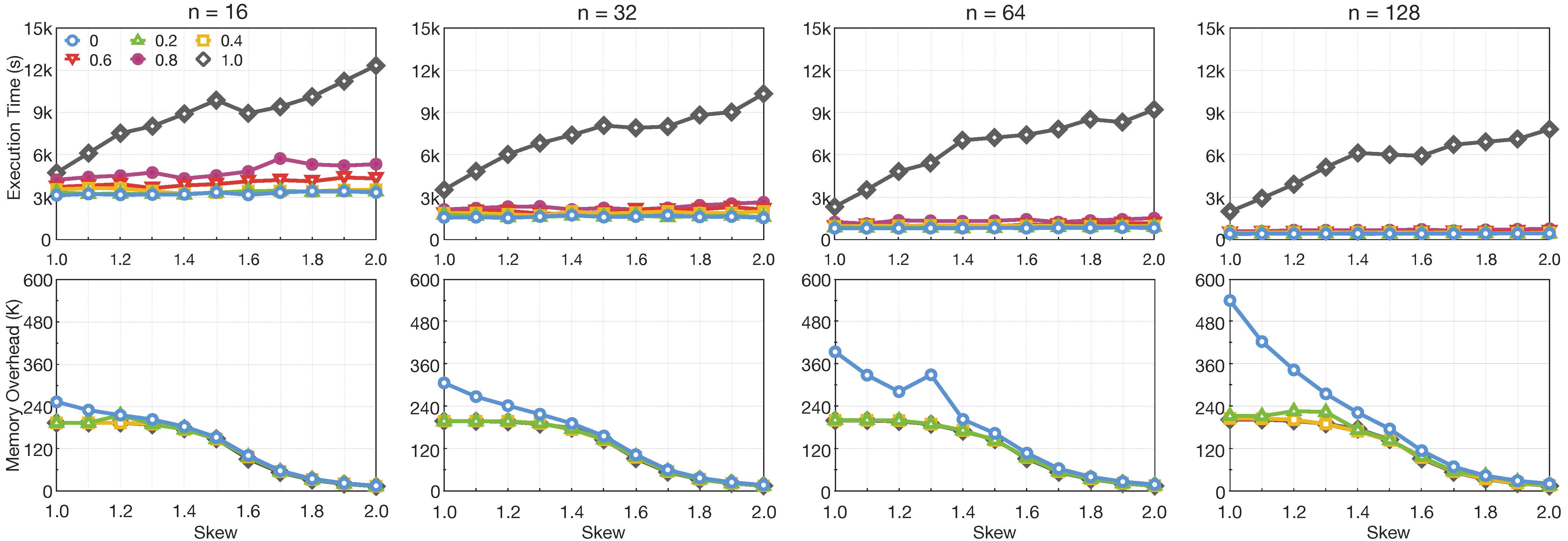}
    \vspace{-1em}
  \caption{Execution time and memory overhead as a function of skew $z$ with different time decaying factor $\alpha$. The results are collected on different number (16/32/64/128) of workers.}
    \vspace{-1.5em}
  \label{fig:alpha}
\end{figure*}

\textbf{Setting Decaying Factor $\alpha$}\quad Our goal is to find an appropriate $\alpha$ so that more stream data can be processed. 
Figure~\ref{fig:alpha} shows the impact of $\alpha$ value, ranging from $0$ to $1$, with different number of workers and skew.

Overall, the large $\alpha$ value can lead to the long execution time. Note that, when $\alpha$ is with $1$, this shows the special case that does not consider the time-evolving feature. We thus can see that the execution time grows significantly (up to 12.14x compared to $\alpha$ of $0.2$) as the skew increases. When $\alpha$ is with $0$, all previous data for each update will be abandoned, although the execution time has a relatively-low level. An amount of memory overhead will be incurred, especially for low skew stream data (with 2.65x compared to $\alpha$ of $0.2$). The reason is that abandoning previous data may mis-lead to many false non-hot keys that are supposed to be hot. Among all possible values, we can see that $\alpha$ with $0.2$ has the best effect on load balance and memory overhead for many cases with different workers and skew.

\begin{figure*}[ht]
  \centering
  \includegraphics[width=7.0in]{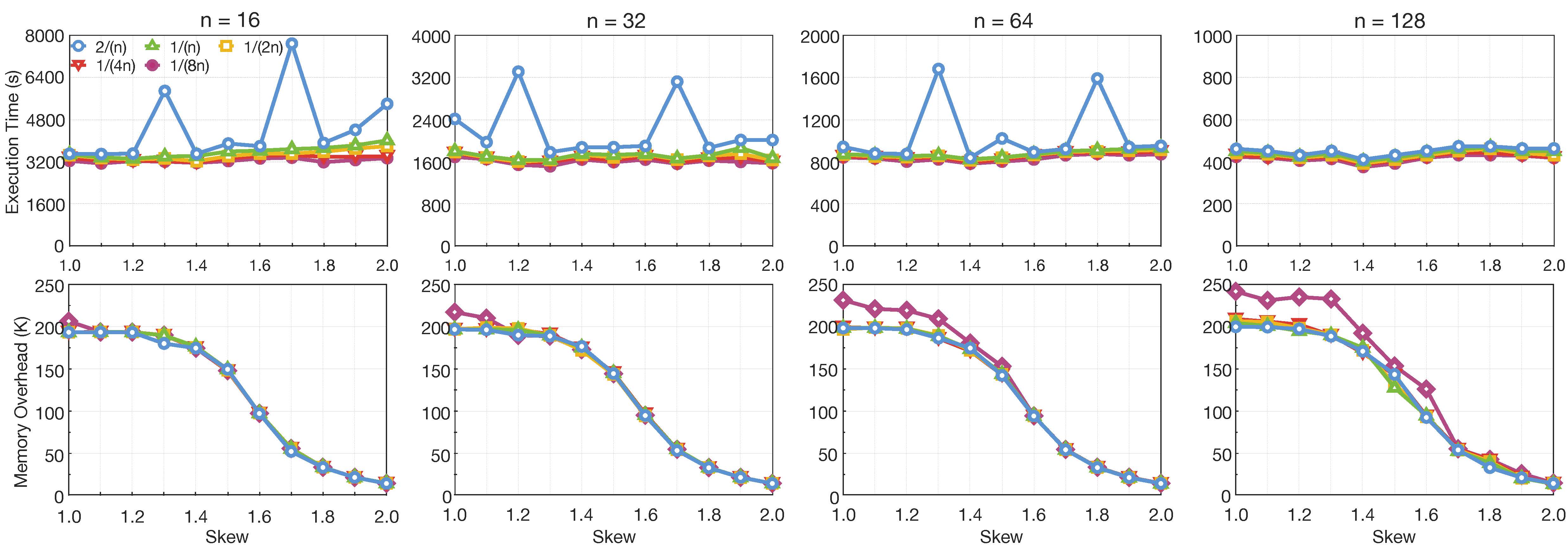}
    \vspace{-1em}
  \caption{Execution time and memory overhead as a function of skew with different hot key thresholds $\theta$. The results are collected with different number (16/32/64/128) of workers.}
    \vspace{-1.5em}
  \label{fig:theta}
\end{figure*}

\textbf{Setting Hot Key Threshold $\theta$}\quad As discussed in the previous study~\cite{nasir2016two}, if $\theta$ is greater than $2/n$ where $n$ is the number of workers, the DSPE can definitely suffer from load imbalance. If it is less than $1/5n$, the probability of load imbalance generated by PKG is bounded by $1$ - $1/n$ at least. An appropriate threshold $\theta$ often lies in the range of from $2/n$ down to $1/8n$. Figure~\ref{fig:theta} shows the potential impact with different $\theta$ thresholds.

In theory, the small threshold often results in the better load balance. The large threshold often results in the lower memory overhead. However, in practice, we can find in Figure~\ref{fig:theta} that significant load imbalance occurs only in the case of $\theta = 2/n$. For other thresholds, the result has almost no difference especially for the large number of workers. As for the memory overhead, we find that memory overhead has little change as $\theta$ is changing. We conservatively choose a threshold as $1/4n$ for two reasons.

First, its execution time is similar to the threshold of $1/8n$ which reflects the similar effect of load balancing. Second, as for memory overhead, it is almost no difference compared to the threshold of $2/n$. However, the threshold with $1/8n$ produces more memory overhead for the large number of workers and low skewed data. A compromised threshold with $1/4n$ can provide the reasonable results on both load balance and memory overhead. 

\begin{figure}[t]
  \centering
  \includegraphics[width=3.3in]{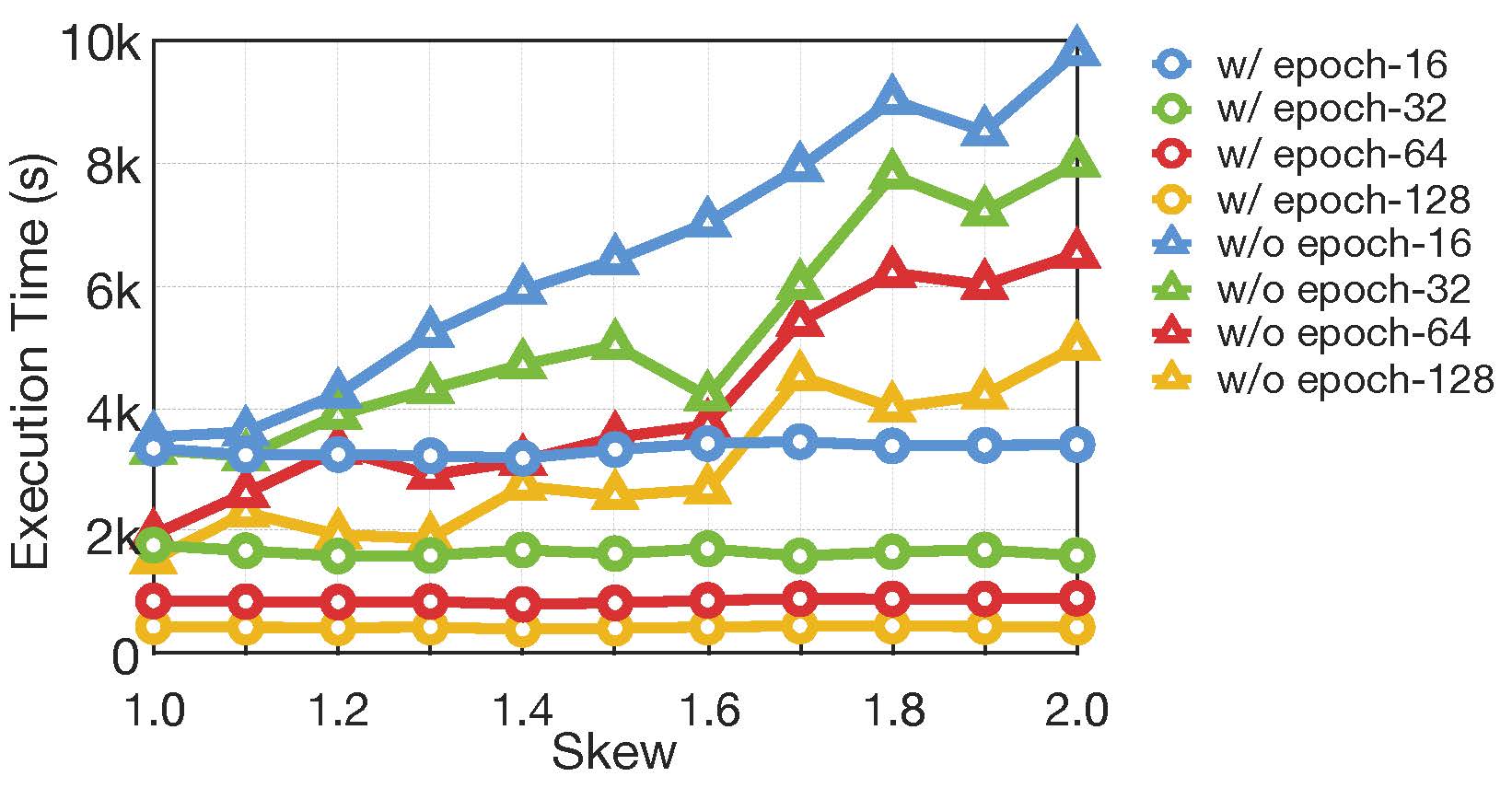}
    \vspace{-1em}
  \caption{Execution time of FISH with and without our epoch-based hot key identification, denoted as w/(o) epoch. The results are collected on different number (16/32/64/128) of workers.}
    \vspace{-1em}
  \label{fig:BOX_I}
\end{figure}

\subsection{RQ3: Breakdown}
We next break down the effectiveness of FISH, including recent hot-key identification, hot-key classification and heuristic worker assignment.

\textbf{Effectiveness of Epoch-based Hot Key Identification}
Figure~\ref{fig:BOX_I} shows the effectiveness of our epoch-based hot key identification compared to the entire lifetime counting-based approach in D-C and W-C. We can see that the execution time has been greatly improved. Especially when the number of workers and the skew increase, the effect becomes more pronounced (up to 11.91x). The main reason accounting for this is that hot key identification in D-C and W-C may potentially lead to inaccurate hot-keys. They monitor the entire lifetime of all keys, thereby resulting in the situation that the most recent hot keys are difficult to capture. This can thus lead to load imbalance among workers. More workers and larger skew can further aggravate the problem of load balance.

\begin{figure}[t]
  \centering
  \includegraphics[width=3.3in]{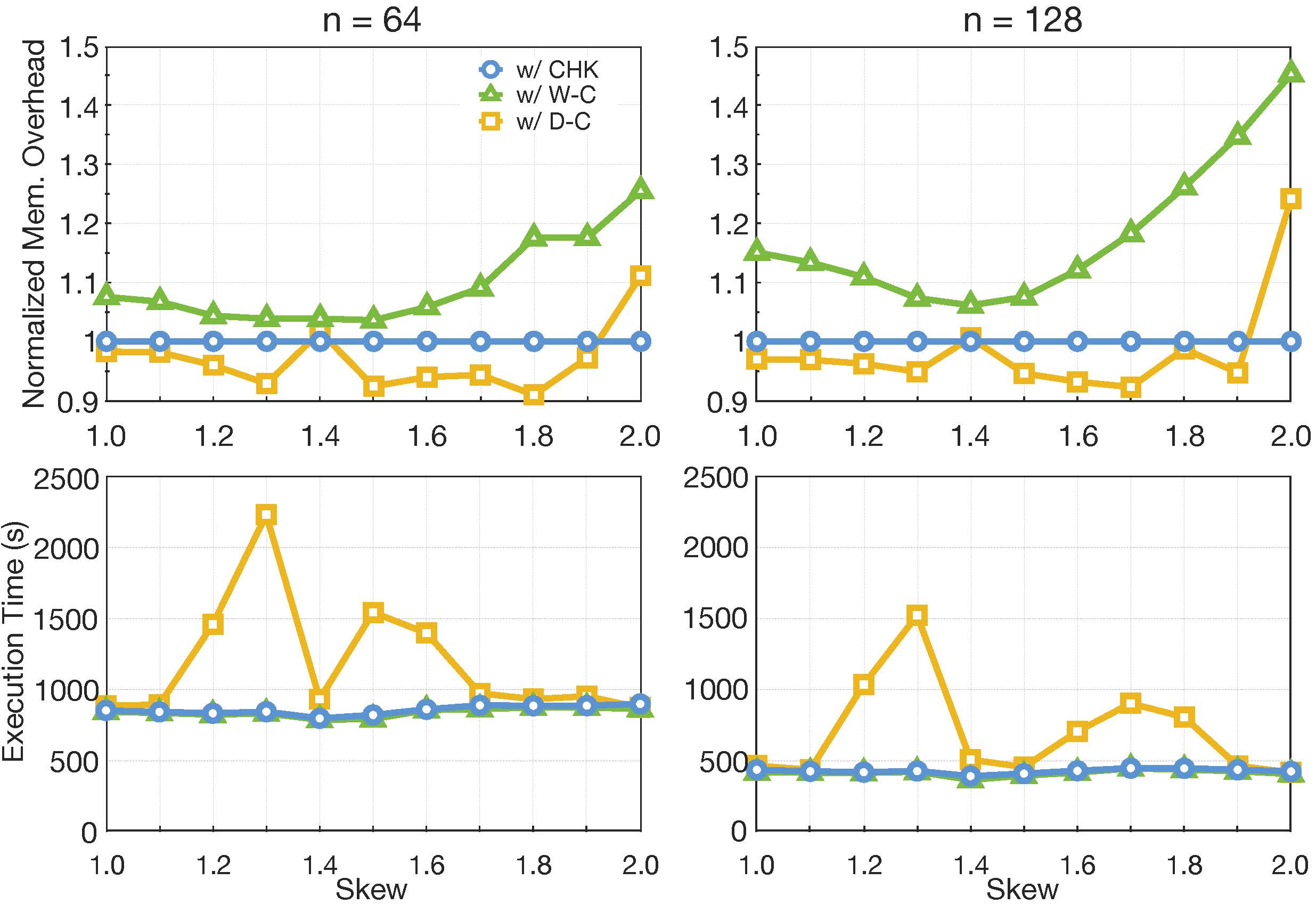}
    \vspace{-1em}
  \caption{Memory overhead and execution time of using different strategies in D-C and W-C against our CHK. The memory overhead results are normalized to CHK. The results are collected on different  number (64/128) of workers.}
    \vspace{-1em}
  \label{fig:MR}
\end{figure}

\textbf{Effectiveness of Hot Key Classification} Figure~\ref{fig:MR} illustrates the memory overhead of FISH with and without CHK. FISH without using CHK includes two cases of hot-key processing approaches that are used in W-C (written as w/ W-C) and D-C (w/ D-C), respectively.

As shown in Figure~\ref{fig:MR}, we can see that CHK can greatly reduce the memory overhead in comparison to the one of W-C. This benefit can be more significant as the number of workers increases. Compared to the method used in W-C, FISH can save up to $25.23\%$ and $45.34\%$ of memory costs for $64$ and $128$ workers respectively. Although the method used in D-C has the less memory overhead than CHK in some cases, it may suffer from longer execution time and more serious load imbalance problems than CHK. Due to the skew distribution of keys, the frequency of hot keys usually varies dramatically. Simply treating all hot keys equally often results in load imbalance (for D-C) or unnecessary memory overhead (for W-C). 
\begin{figure}[t]
  \centering
  \includegraphics[width=3.1in]{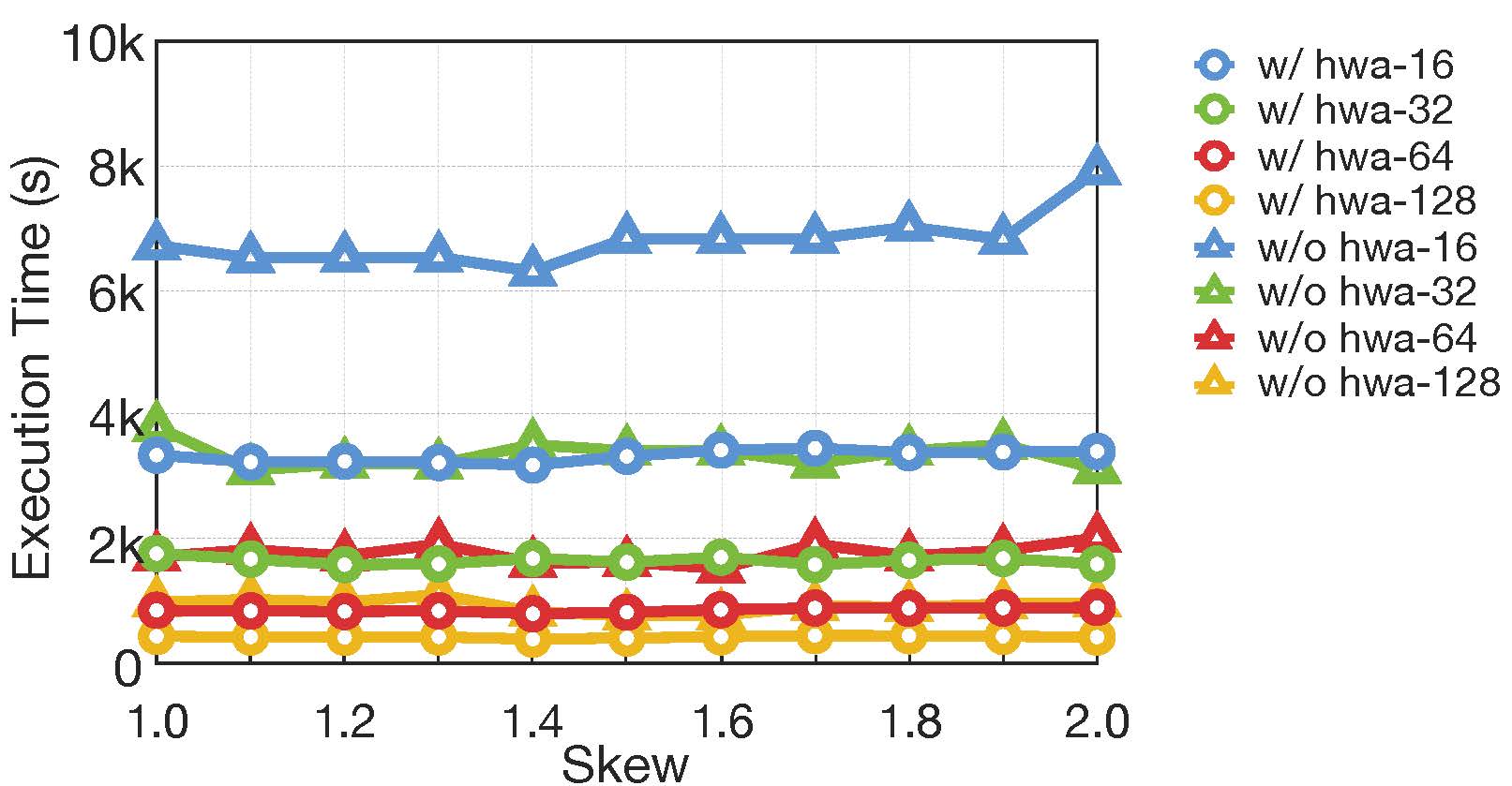}
  \vspace{-1em}
  \caption{Execution time of FISH with and without heuristic worker assignment. We collect results with different number (16/32/64/128) of workers.}
  \vspace{-1em}
  \label{fig:Hen}
\end{figure}

\textbf{Effectiveness of Heuristic Worker Assignment}\quad In order to verify the effectiveness of heuristic worker assignment (hwa), we assume that half of the worker's processing capability is twice than the others.

Figure~\ref{fig:Hen} plots the results. We can see that FISH can provide up to 2.61x improvement on the execution time compared to the traditional worker assignment in previous studies~\cite{nasir2015power,nasir2016two} which assigns the keys according to the amount of worker's load. The main reason accounting for this is that simply ensuring each worker has the same number of tuples in the final state may assign a busy worker for a tuple in some time interval, particularly true for the situation where workers have different processing capacity. In contrast, our approach is able to cope with scenarios where workers are heterogeneous and dynamically changing by inferring the status of workers.

\begin{figure}[t]
  \centering
  \includegraphics[width=3.3in]{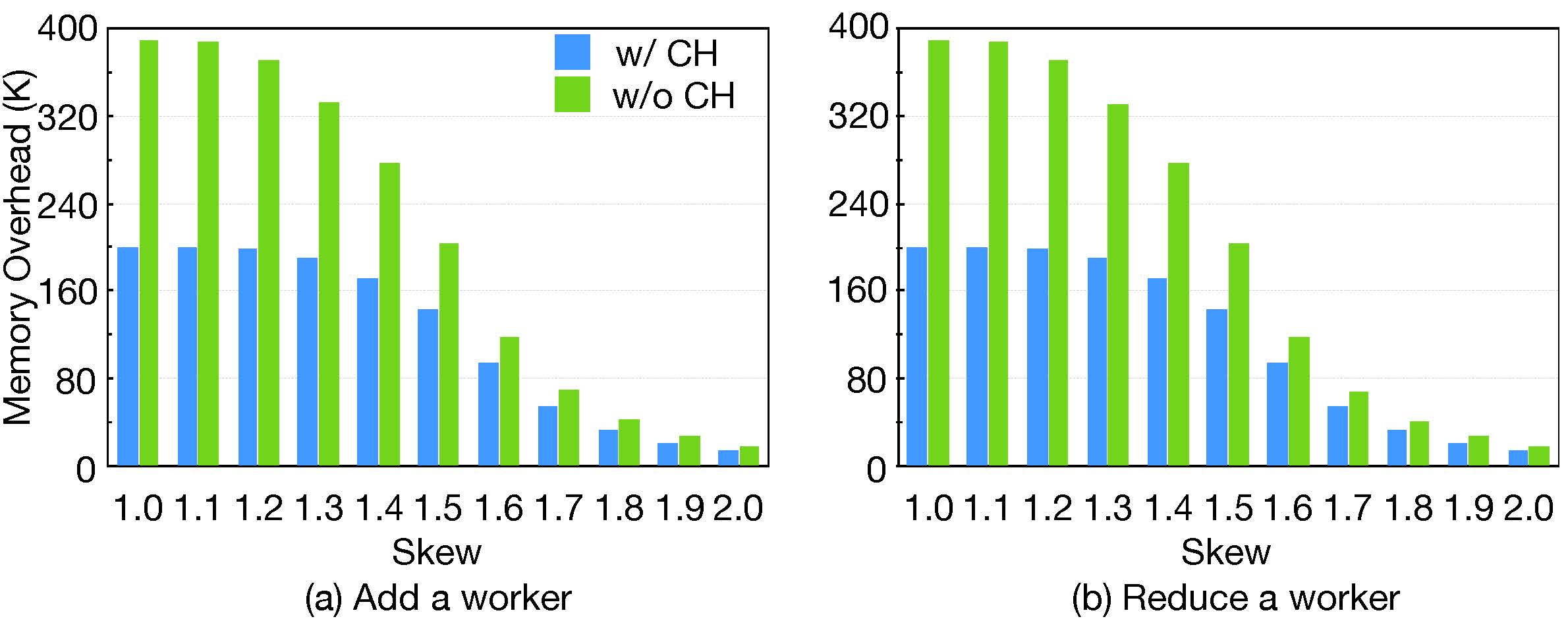}
  \vspace{-1em}
  \caption{Memory overhead of FISH with/without consistent hashing (CH) for dynamic change of workers. (a) Add a worker in a half execution of task; (b) Reduce a worker in a half execution of task. }
  \vspace{-1em}
  \label{fig:CH_J}
\end{figure}

\subsection{RQ4: Effectiveness of Consistent Hashing}
In order to investigate the effectiveness of consistent hashing (CH), we create the dynamic scenario by randomly adding or removing a worker instance during the processing.

Figure~\ref{fig:CH_J} illustrates the memory overhead of FISH with and without CH with different skewed stream data. As we can see, for stream data with low skew, FISH without CH almost has memory overhead twice than FISH with CH no matter the workers are increased or decreased. This is because that the previous key and worker mappings rely heavily on the number of workers. The variation of worker number just means that almost all possible mappings need to be changed, leading to twice memory overhead.
Stream processing on highly-skewed dataset has less increase of memory overhead. The reason for this is that the hot keys for stream dataset with high skewness need to be re-mapped to new workers. Considering a part of new workers have already reserved the corresponding data of these hot keys. As a result, this can save an amount of memory overhead so that not too much remapping has occurred when the number of workers is changing.



\subsection{RQ5: Practical Deployment on Apache Storm}
To quantify the impact of FISH, we have integrated it into Apache Storm and deployed it on a cluster with 8 compute nodes, each of which has 20 available ports. We build a DAG topology configured with 32 sources and 128 workers. We compared FISH with state-of-the-art FG, SG, PKG, D-C and W-C grouping schemes.

\begin{figure}[t]
  \centering
  \includegraphics[width=3.3in]{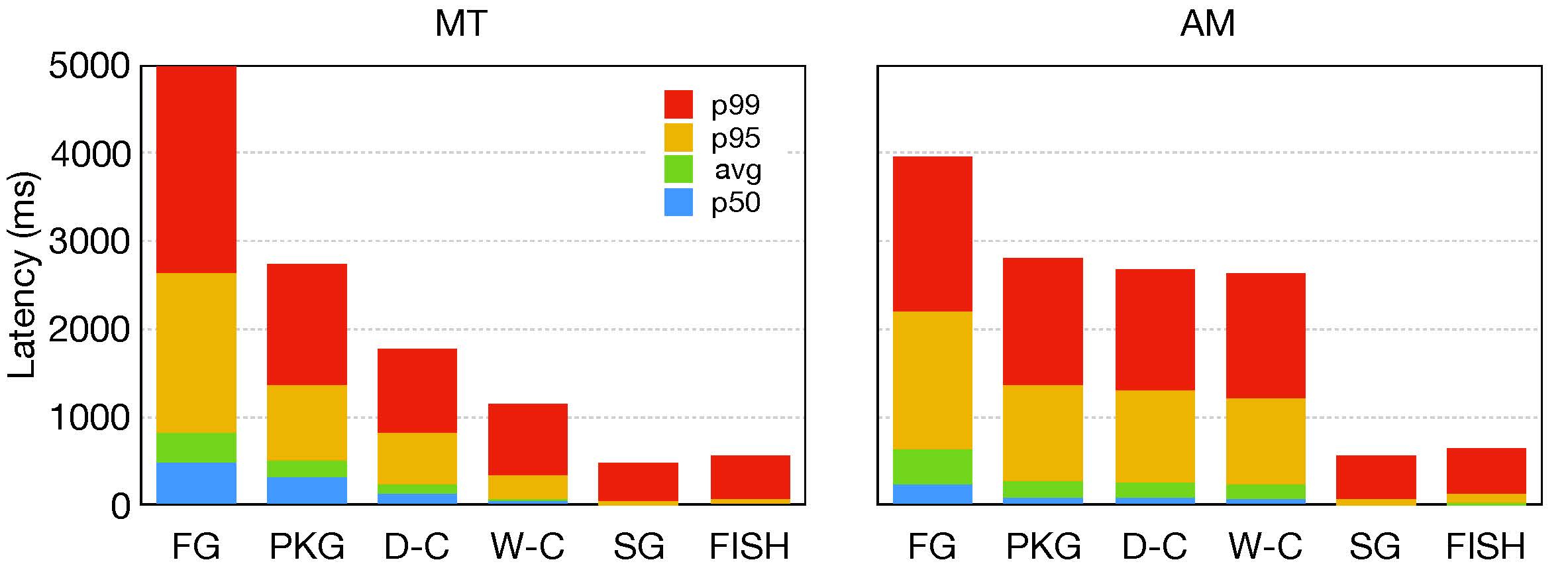}
  \vspace{-1em}
  \caption{The average and percentiles latency by deploying FG, PKG, D-C, W-C, SG, and FISH  on Apache Storm with the MT and AM datasets}
  \vspace{-1em}
  \label{fig:latency}
\end{figure}

\textbf{Latency}\quad Figure~\ref{fig:latency} shows the results regarding end-to-end latency. The plot reports the average latency with the 50th, 95th, and 99th percentiles across all workers, respectively. 
Thanks to the accurate hot key identification and heuristic worker assignment. The 50th (median) and 99th percentiles in FISH have the geometric mean of latency with only 7 and 562 milliseconds (for MT), as well as 9 and 640 milliseconds (for AM), respectively. These results are almost the ideal latency provided by SG. In summary, FISH significantly outperform FG, W-C, D-C, and PKG. FISH can reduce the average and 99th percentile latency of state-of-the-art W-C by 87.12\% and 76.34\%, respectively.

\begin{figure}[t]
  \centering
  \includegraphics[width=3.3in]{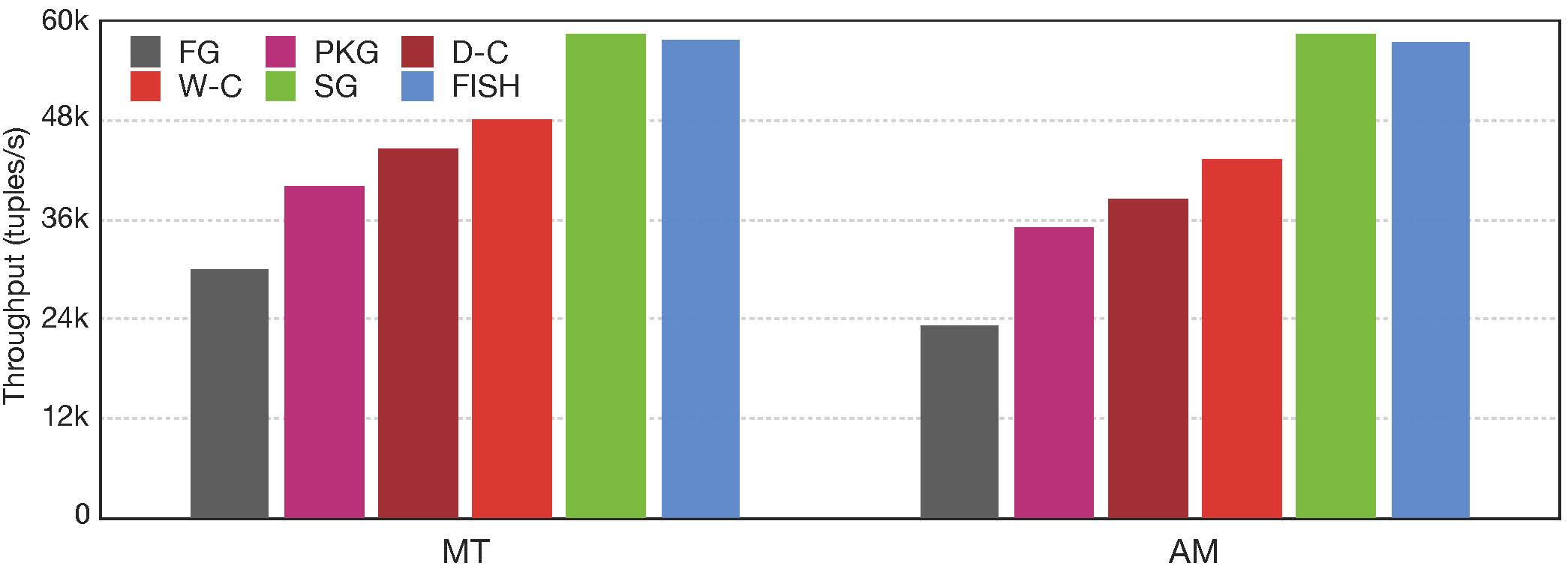}
  \caption{Throughput comparison by deploying FG, PKG, D-C, W-C, SG, and FISH on Apache Storm with the MT and AM datasets}
  \vspace{-1em}
  \label{fig:throughput}
\end{figure}

\textbf{Throughput}\quad Figure~\ref{fig:throughput} shows the results regarding throughput. Overall, FG has the lowest throughput (with 30K tuples/sec for MT and 23K tuples/sec for AM). Compared to FG, PKG involves a considerable improvement. Further, D-C and W-C perform better than PKG, but still have a distance gap for matching the throughput of SG. In comparison, FISH can provide a throughput 1.32 times higher than W-C, and 1.48 times higher than D-C. On the whole, we can observe that FISH can provide the almost ideal throughput close to the one by SG.

\begin{figure}[t]
  \centering
  \includegraphics[width=2.8in]{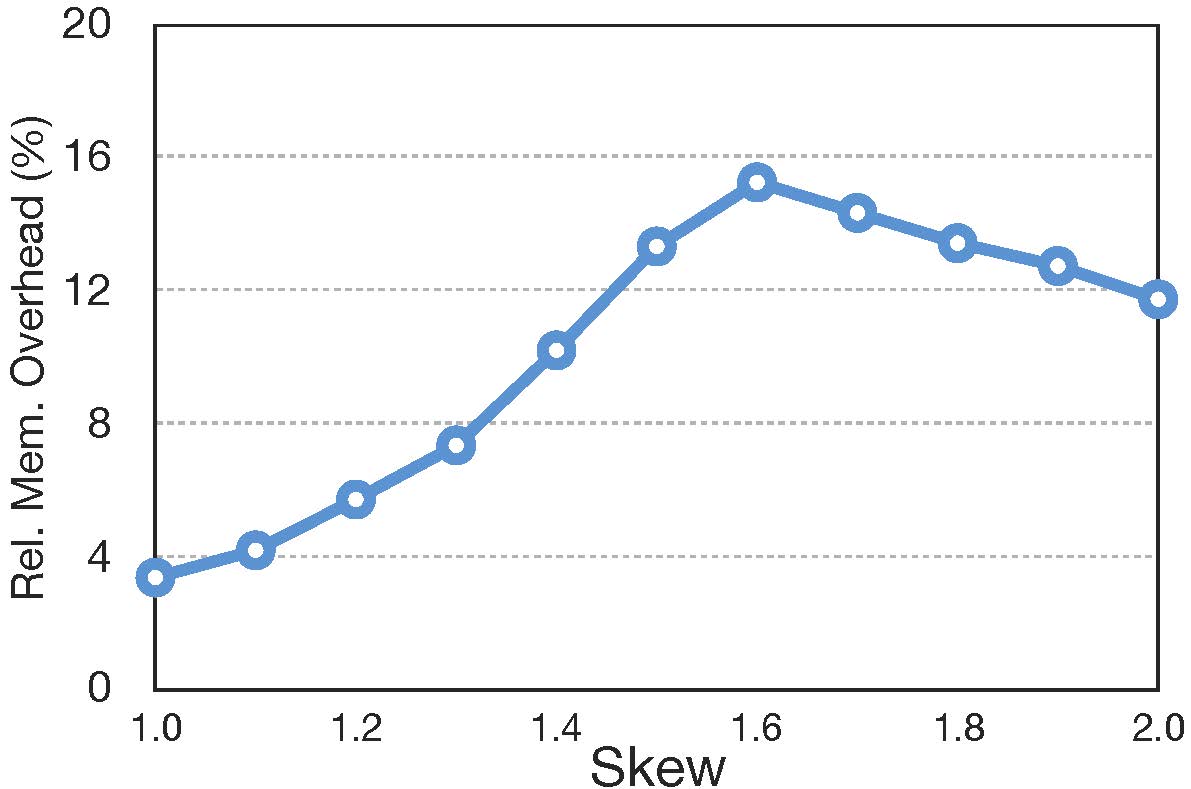}
  \vspace{-1em}
  \caption{Relative memory overhead of FISH against SG}
  \vspace{-1.5em}
  \label{fig:mem}
\end{figure}

\textbf{Memory Overhead}\quad As discussed above, we can see that SG provides the best effect on load balance in terms of latency and throughput. We next investigate the comparative results of memory overhead of FISH in comparison to SG. Figure~\ref{fig:mem} plots the normalized results with different skewness. The baseline is the results with SG.
We can find that, for the skew with 1.0, the memory overhead in FISH can be as low as $3.34\%$ of that in SG. Overall, FISH has significantly less ($<16\%$) memory overhead than SG.


{\bf Summary}\quad According to aforementioned results, we find that FISH is able to technically provide the compelling latency and throughput results of SG grouping scheme for time-evolving stream data at a very small fraction of memory overhead.

\section{Related work}
\label{sec:Related work}
A large number of previous studies~\cite{gedik2014partitioning,rivetti2015efficient,shah2003flux,xing2005dynamic,castro2013integrating,gedik2014elastic,kumbhare2015fault} leverage operator migration for load balance in DSPEs. Once a situation of load imbalance is detected, the system activates a rebalancing routine that moves some keys and their associated states away from an overloaded server.

Flux~\cite{shah2003flux} encapsulates adaptive state partitioning and dataflow routing, migrates operators from the most loaded to the least loaded server. Xing et al.~\cite{xing2005dynamic} present a correlation based load distribution algorithm for dynamic load migration to adapt to changing loads. 
Fernandez et al.~\cite{castro2013integrating} propose an integrated approach for scale-out and failure recovery through checkpointing and migration.
Gedik~\cite{gedik2014partitioning} propose partitioning functions for stream processing systems that employ stateful data parallelism to improve application throughput and control migration cost. 

These rebalance-based approaches usually require setting a number of parameters, such as how often to check for imbalance. These parameters are typically application-specific with different tradeoff situations between imbalance and rebalance cost. Further, each sub-stream needs to maintain a routing table that maps the key to each PEIs with prohibitive memory overhead. Also, modifying the routing table introduces additional consistency checking across all sub-streams~\cite{nasir2015power}. In contrast, we consider operators replication that allows the key can be processed by multiple workers and show it is sufficient to balance the load without active monitoring of the load imbalance.

A wide spectrum of studies attempt to consider operator replication to prevent load imbalance~\cite{nasir2015power,nasir2016two,nasir2017load}. They allow that each key can be processed by multiple workers. POTC~\cite{nasir2015power} based on the “power of two choices”~\cite{azar1999balanced} which associates each key to two possible operator instances, and selects the minimum load of the two whenever a tuple for a given key must be processed. Nasir et al.~\cite{nasir2016two} propose a lightweight streaming grouping scheme which is based on the SpaceSaving~\cite{karger2004simple} algorithm and does not require training or monitoring to detect the heavy hitters. CG~\cite{nasir2017load} studied the load balancing problem for streaming engines running in a heterogeneous cluster. Our specialized approach differs from these replication-based approaches with the following significant innovation: 1) We first consider the feature of time-evolving stream data and investigate real-time load balance within some time interval; 2) We present a novel heuristic method to assess the state information of remote workers for efficient worker assignment.

There also involves much effort put into operator placement, which ensures load balance by exploiting computational resources. Xing et al.~\cite{xing2006providing} propose a correlation-based algorithm that strives to minimize operator movement overhead and support more resilient operator placement.~\cite{aniello2013adaptive} deploys a topology via using both online and offline analyzing methods under the minimal network communication. Eidenbenz et al~\cite{eidenbenz2016task} analyze the task allocation problem and propose an approximation algorithm to exploit optimal solution. In contrast to these studies with resource partition, our approach makes workload partition for load balance. Note that our approach is compatible with an integration of this type of approach with a hybrid partition, which can be interesting future work for achieving load balance with minimum computational resources.

\section{Conclusion}\label{sec:Conclusion}
In this work, we investigate the load balance problem for time-evolving stream processing with a large scale deployment. Our key innovation comes from two major technical advances. First, we present an epoch-based approach to identify recent hot keys efficiently by intra-epoch frequency counting and inter-epoch hotness decaying. Second, based on the similarity of operations in streaming processing, we further propose a heuristic approach to infer the state information of remote workers to make the efficient worker assignment. We evaluate our approach on a cluster of 128 nodes with both synthetic and real-world datasets. Our practical deployment on Apache Storm demonstrates that FISH significantly outperforms state-of-the-art with the average and 99th percentile latency reduction by up to 87.12\% and 76.34\% (vs. W-Choices), and 96.66\% memory consumption reduction (vs. Shuffle Grouping).

\ifCLASSOPTIONcaptionsoff
  \newpage
\fi



%

\bibliographystyle{IEEEtran}
\bibliography{mybibfile}
\end{document}